\documentclass[aps,preprint,nofootinbib,preprintnumbers,eqsecnum]{revtex4-1}

\usepackage[usenames, dvipsnames]{color}

\newcommand{\HL}[1]{{\textcolor{magenta}{#1}}}

\usepackage[
	  pagebackref=false,
	  colorlinks=true,
      linkcolor=blue,
      urlcolor=blue,
      filecolor=black,
      citecolor=red,
      pdfstartview=FitV,
      pdftitle={},
        pdfauthor={},
        pdfsubject={},
        pdfkeywords={},
        pdfpagemode=None,
        bookmarksopen=true
      ]{hyperref}

\usepackage[normalem]{ulem}
\usepackage{amsmath}
\usepackage{enumerate}
\usepackage{amsfonts}
\usepackage{epsfig}
\usepackage{mathbbol}

\def\Tr{\mathop{\rm Tr}}
\def\tr{\mathop{\rm tr}}

\newcommand\field[1]{{\ensuremath{\mathbb{{#1}}}}}

\newcommand\vev[1]{{\ensuremath{\left\langle{#1}\right\rangle}}}

\newcommand\ket[1]{\ensuremath{\lvert{#1}\rangle}}
\newcommand\bra[1]{\ensuremath{\langle{#1}\rvert}}

\newcommand{\NN}{\field{N}}

\newcommand{\be}{\begin{equation}}
\newcommand{\ee}{\end{equation}}
\newcommand{\bea}{\begin{eqnarray}}
\newcommand{\eea}{\end{eqnarray}}
\newcommand{\bega}{\begin{gather}}
\newcommand{\eega}{\end{gather}}

\newcommand{\bi}{\begin{itemize}}
\newcommand{\ei}{\end{itemize}}
\newcommand{\ben}{\begin{enumerate}}
\newcommand{\een}{\end{enumerate}}
\newcommand{\bca}{\begin{cases}}
\newcommand{\eca}{\end{cases}}
\newcommand{\bln}{\begin{align}}
\newcommand{\eln}{\end{align}}
\newcommand{\bst}{\begin{split}}
\newcommand{\est}{\end{split}}
\def\ie{\begin{equation}\begin{aligned}}
\def\fe{\end{aligned}\end{equation}}
\newcommand{\bma}{\le(\begin{matrix}}
\newcommand{\ema}{\end{matrix}\ri)}
\newcommand{\bwt}{\begin{widetext}}
\newcommand{\ewt}{\end{widetext}}

\newcommand\al{{\alpha}}
\def\b{{\beta}}

\newcommand\ga{{\ensuremath{{\gamma}}}}

\newcommand\de{{\ensuremath{{\delta}}}}

\newcommand\da{{\dagger}}

\newcommand\Lra{{\Longrightarrow}}
\newcommand\ov{\over}

\def\le{\left}
\def\ri{\right}

\newcommand\sG{{\ensuremath{{\mathcal G}}}}
\newcommand\sH{{\ensuremath{{\mathcal H}}}}

\newcommand\sO{{\ensuremath{{\mathcal O}}}}
\newcommand\sP{{\ensuremath{{\mathcal P}}}}

\newcommand\sR{{\mathcal R}}

\pdfoutput=1

\usepackage{tensor}
\usepackage{braket}
\usepackage{comment}

\newcommand{\coeff}{{|c_{\al}^{\;\b} (t)|^2 }}

\newcommand{\bid}{{\mathbf 1}}
\newcommand{\upa}{{\uparrow}}
\newcommand{\doa}{{\downarrow}}

\newcommand{\seq}{{s_{\rm eq}}}
\newcommand{\rgd}{{random void distribution}}
\newcommand{\slcg}{{sharp light-cone growth}}

\def\XXint#1#2#3{{\setbox0=\hbox{$#1{#2#3}{\int}$}
     \vcenter{\hbox{$#2#3$}}\kern-.5\wd0}}

\begin{document}

\title{Void formation in operator growth, entanglement, and unitarity}

\preprint{MIT-CTP/5169}

\author{Hong Liu and Shreya Vardhan}
\affiliation{Center for Theoretical Physics, \\
Massachusetts
Institute of Technology,
Cambridge, MA 02139 }

\begin{abstract}

 \noindent The structure of the Heisenberg evolution of operators plays a key role in explaining diverse processes in quantum many-body systems. In this paper, we discuss a new universal feature of operator evolution: an operator can develop a void during its evolution, where its nontrivial parts become separated by a region of identity operators.  Such processes are present in both integrable and chaotic systems, and are required by unitarity.  We show that void formation has important implications for unitarity of entanglement growth and generation of mutual information and multipartite entanglement. 
%As an application, we argue that operators that make up the density operator of a black hole can ``jump'' outside the black hole after the Page time, providing the underlying physical mechanism  for  a recent semi-classical prescription for the resolution of a  black hole information loss puzzle. 
We study explicitly the probability distributions of void formation in a number 
of unitary circuit models, and conjecture that in a quantum chaotic system the distribution is given by the one we find in random unitary circuits, which we refer to as the 
random void distribution. We also show that random unitary circuits lead to the same pattern of entanglement growth for multiple intervals as in $(1+1)$-dimensional holographic CFTs after a global quench, which can be used to argue that the random void distribution leads to maximal entanglement growth.
% and suggests that it underlies the time-evolution of holographic systems.}

\end{abstract}

\today

\maketitle

\tableofcontents

\section{Introduction}
The Heisenberg evolution of operators in a quantum many-body system is in general extremely complicated. But during the last decade, 
remarkable universalities have been found, such as ballistic growth of operators \cite{Roberts} and growth of operator entanglement \cite{prosen1, prosen2, dubail, JHN}. Such universal properties have played an important role in diverse problems like
scrambling of quantum information, quantum many-body chaos, and entanglement growth during thermalization (see e.g.~\cite{Shenker1,abanin}).

\begin{figure}[!h]
\begin{center}
\includegraphics[width=11cm]{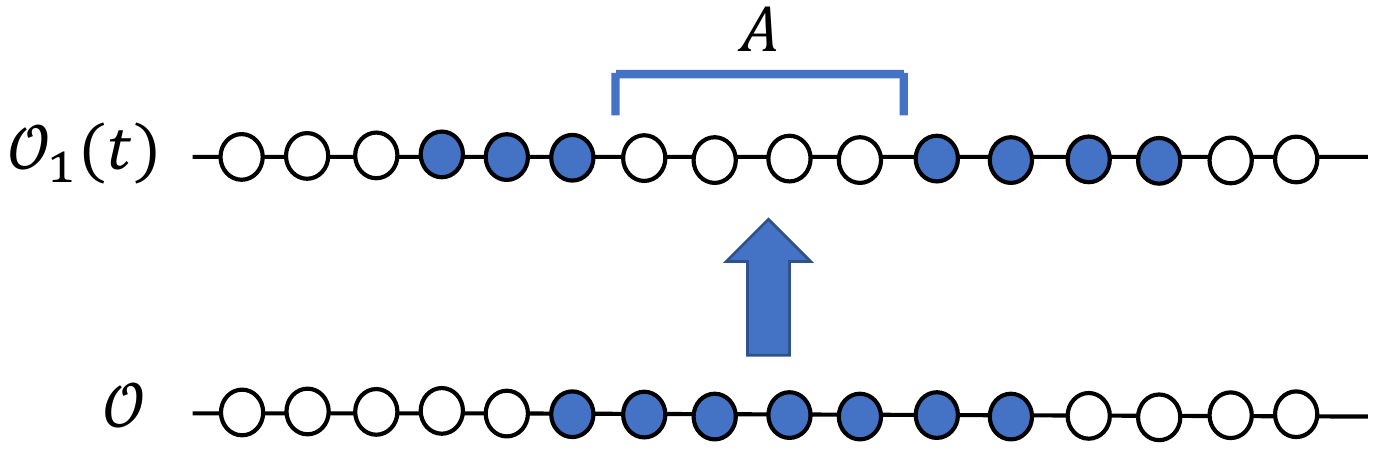} 
\caption{An operator $\sO$ in a many-body system can develop a ``void'' under time evolution, where nontrivial parts of $\sO_1 (t)$ (introduced in~\eqref{yhe}) are separated by a region of identity operators.  In the example shown above in one spatial dimension, the operator has nontrivial single-site operators at blue sites, and the identity at white sites.}
\label{fig:gap}
\end{center}
\end{figure}

In this paper, we discuss another universal feature of operator evolution: an operator can develop a void during its evolution.  
More explicitly, considering a spatial region $A$ within the ``lightcone'' of an initial operator $\sO$, 
we can decompose its time evolution $\sO (t)$  as 
\be \label{yhe}
\sO (t) =\sO_1 (t) + \sO_2 (t) , \qquad  \sO_1 (t)= \tilde \sO_{\bar A} \otimes \bid_A
\ee
where $\bid_A$ denotes the identity operator in $A$, $\tilde \sO_{\bar A}$ is some operator  
in $\bar A$ (the complement of $A$)\footnote{In the more detailed definition of a void that we give in section \ref{sec:unit}, we will only include the part of $\tilde{\mathcal{O}}_{\bar{A}}$ which is orthogonal to the identity operator when projected onto any disconnected region of $\bar{A}$.}, and $\sO_2 (t)$ 
is an operator whose projection  onto $A$ is orthogonal to $\bid_A$. Here we assume that the system has a finite-dimensional Hilbert space and tensor product structure associated with spatial regions. 
See Fig.~\ref{fig:gap} for an illustration. Given the space of all operators is a Hilbert space, we can also associate a weight or ``probability'' for $\sO (t)$ to develop a void in region $A$
\be 
P_{\sO}^{(A)} (t) = {\Tr \le(\sO_1^\da (t) \sO_1 (t) \ri) \ov \Tr \le(\sO^\da (t) \sO (t) \ri) }  \ .
\ee

\begin{figure}[!h]
\begin{center}
\includegraphics[width=15cm]{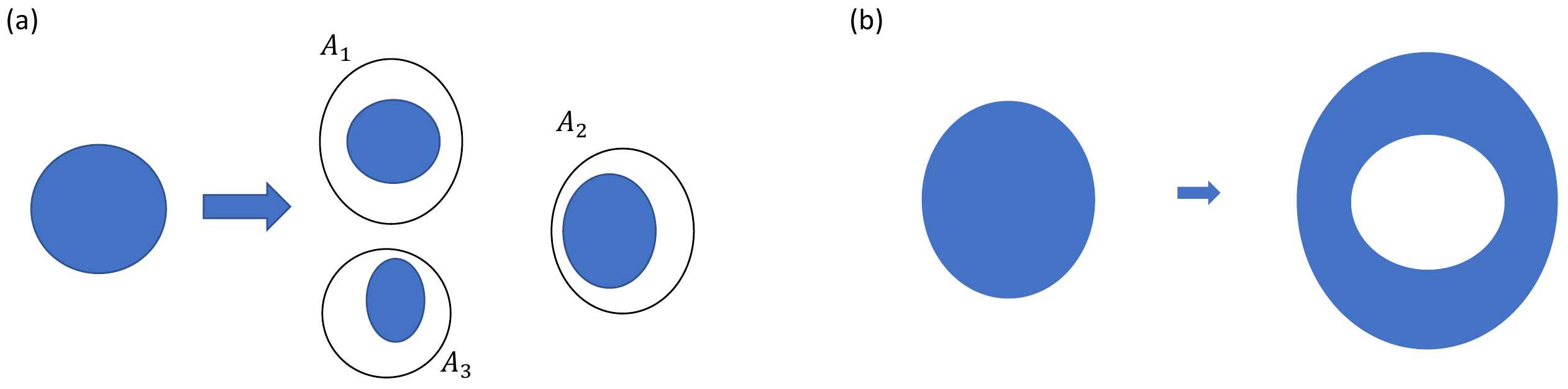} 
\caption{{In more than one spatial dimension, void formation can either break up a single operator into disconnected parts, as in (a), or form a hole in an operator, as in (b). In the figure, shaded regions indicate
where the operator has nontrivial support. A process like (a) contributes to multipartite
entanglement between the three disjoint regions $A_1$, $A_2$, and $A_3$. 
}}
\label{fig:mul}
\end{center}
\end{figure}

Below we will refer to the presence of $\sO_1 (t)$ in $\sO(t)$ as void formation. Void formation is present in both integrable and chaotic systems, and is required by unitarity. We will show that it has important implications for unitarity of entanglement growth. 
For example, evolving from an initial product state, it is the contribution from $\sO_1 (t)$ that ensures  $S_A (t) = S_{\bar A} (t)$, where $S_A$ is the entanglement entropy for region $A$. Further, we will show that void formation is responsible for generating mutual information and multipartite entanglement among disjoint regions during the evolution of a system, as we illustrate using a cartoon picture in Fig.~\ref{fig:mul}. 
We will also derive a number of general constraints on probability distributions of void formation from unitarity, which are applicable to 
both chaotic and integrable systems.

To develop intuition for probability distributions of void formation, we study three types of unitary circuit models in one spatial dimension: (i) 
the random unitary model of~\cite{nahum1,frank,nahum2}, which can be considered a minimal model for quantum chaotic systems; (ii)  a ``free propagating'' model~\cite{hong_mark} in which entanglement 
can only be spread, but not created, which may thus be considered a proxy for free theories; 
(iii)  a circuit built from perfect tensors~\cite{qi}, which may be considered a model for interacting integrable systems. In the free-propagating and perfect tensor models, patterns of void formation depend sensitively on the initial operator. In particular, in the perfect tensor circuit model, void formation exhibits a fractal structure.

In the random unitary circuit, we find at sufficiently late times %the probability for a generic operator $\sO$  to develop a void in some region $A$ is given at sufficiently late times by %which can include disconnected parts, is given by
\be \label{pgd}
P^{(A)}_\sO = {1 \ov d_A^2} = e^{- 2 S_{\rm eq}^{(A)}} 
\ee
where $d_A$ is the dimension of local Hilbert space in region $A$. After the second equality, we have written the expression in a form which is generalizable to continuum systems, with $S_{\rm eq}^{(A)}$ the equilibrium entropy of $A$. 
It is natural to conjecture that~\eqref{pgd}, to which we will refer as the \rgd\ from now on, holds for generic operators in general chaotic systems at sufficiently late times. 
While~\eqref{pgd} is very small for a macroscopic region $A$, for certain processes the number of contributing operators can be exponentially large, leading to significant physical effects.

As an illustration, we show that together with the assumption of sharp light-cone growth of operators, the \rgd~\eqref{pgd} fully determines the second Renyi entropy of an arbitrary number of disjoint intervals in random unitary circuits in the limit of  
large one-site Hilbert space dimension. Furthermore, surprisingly,  the resulting expression
coincides exactly with the von Neumann entropy  after a global quench in $(1+1)$-dimensional holographic systems.
On the one hand, this indicates that the \rgd~\eqref{pgd} may underlie operator evolution in holographic systems. 
On the other hand,  in the light of the fact that the holographic expression maximizes the evolution of entanglement entropy in all $(1+1)$-dimensional systems~\cite{hong_mark}, we are led to conclude that together with sharp light-cone growth, the \rgd\ maximizes entanglement growth.

The plan of this paper is as follows. In Sec.~\ref{sec:unit}, after describing in detail our general set-up, we discuss implications of void formation for unitarity of entanglement growth and generation of mutual information and multi-partite entanglement, as well as constraints on void formation from requiring unitarity. 
In Sec.~\ref{sec:rnf}, we discuss the random circuit model, derive the \rgd~\eqref{pgd}, and discuss its implications. 
In Sec.~\ref{sec:circ}, we discuss void formation in the free propagation and perfect tensor models. 
We conclude in Sec.~\ref{sec:conc} with future directions. We have included a number of Appendices for technical details.

\section{Void formation and implications} \label{sec:unit}

In this section, we first describe our general setup, and then derive some simple constraints from unitarity on void formation during Heisenberg evolution.

\subsection{Setup} \label{sec:set}

For convenience, we will consider a one-dimensional lattice system with a finite-dimensional Hilbert space at each site. The discussion generalizes immediately to higher dimensions. We comment on generalizations to systems with an infinite local Hilbert space in the discussion section, Sec.~\ref{sec:conc}. 

The Hilbert space at a site $i$ will be denoted as $\sH_i$,  and is taken to have dimension $q$. The full 
Hilbert space is $\sH = \otimes \sH_i$, and has dimension $q^L$, where $L \to \infty$ is the system size. 
Operators at a single site form a Hilbert space of dimension $q^2$, which will be denoted as $\sG_i$. 
Operators of the full system
form a Hilbert space $\sG = \otimes_i \sG_i$ of dimension $q^{2 L}$. We will use $O_a^i, a =0,1, \cdots q^2-1$, to denote an orthonormal basis of $\sG_i$ which is normalized as 
\be \label{jen}
\tr ((O_a^i)^\da O_b^i)  = q \de_{ab} , \qquad O_0^i = \bid_i
\ee
where $\bid_i$ is the identity operator of $\sH_i$, and there is no summation over $i$ in the above equation. Orthogonality with $O_0^i$ implies $O_{a}^i, a =1, \cdots q^2-1$, are all traceless. A convenient choice of basis which we will use throughout the paper is 
\be \label{obd}
O_a = X^{s_1} Z^{s_2}, \qquad s_1,s_2 = 0,1, \cdots q-1
\ee
where $X$ and $Z$ are respectively 
 the shift and clock matrices\footnote{Explicitly, $Z = \sum_{k=0}^{q-1}e^{2\pi i k/q}\ket{k}\bra{k}$ and $X = \sum_{k=0}^{q-1}\ket{k+1}\bra{k}$, where addition is defined mod $q$.}. 
 An orthonormal basis for $\sG$, which will be denoted as 
$\sO_\al, \al =0,1, \cdots q^{2L}-1$,  can be obtained from tensor products of $\{O_a^i\}$. These basis operators satisfy
\be \label{hne}
%\sO_\al^\da = \sO_\al , \qquad 
\Tr \sO_\al^\da \sO_\b = \de_{\al \b} ~ q^L  \ .
\ee
$\sO_0$ is the identity operator for the full Hilbert space $\sH$, and all other $\sO_\al$'s are traceless. 

Under time evolution, 
\be \label{evop}
\sO_\al (t) = U^\da (t) \sO_\al U(t) =  \sum_\b c_\al^\b (t) \sO_\b
\ee
where $U(t)$ denotes the evolution operator. From unitarity of $U(t)$, 
\be 
\sum_\b |c_\al^\b  (t) |^2  =1  \ .
\ee
We can interpret $| c_\al^\b  (t) |^2$ as the probability of operator $\sO_\al$ evolving to $\sO_\b$.  
Systems with a local Hamiltonian have an effective light-cone speed $v_c$ for how fast an operator can grow with time~\cite{LR}, so that $|c_\al^\b  (t) |^2 \approx 0$  for $\sO_{\beta}$ not contained within the light-cone of $\sO_{\alpha}$. 

\begin{figure}[!h]
\begin{center}
\includegraphics[width=10cm]{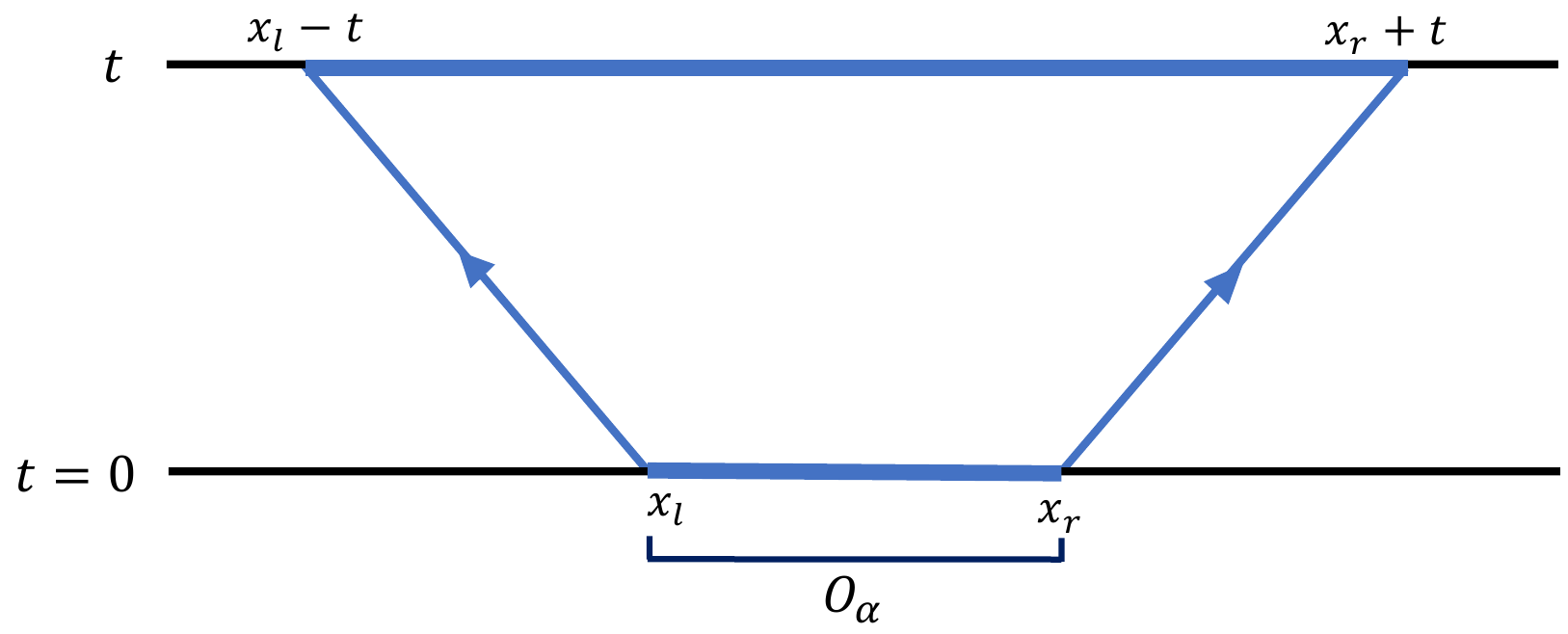} 
\caption{Under time-evolution, the end points of an operator $\sO_\al$ move at the light-cone speed $v_c=1$ in opposite directions. The initial operator with endpoints at $x_l$ and $x_r$ can become a superposition of many final operators at time $t$, each of which have endpoints $x_l-t$ and $x_r+t$.}
\label{fig:lc}
\end{center}
\end{figure}

For the purpose of not obscuring the conceptual picture with technicalities,  throughout this paper, we will consider a particularly simple form of operator evolution in which the end points of an operator of interest move at the light-cone speed 
$v_c$ in opposite directions\footnote{
Examples include the random unitary circuits in the large $q$ limit~\cite{nahum1,frank}, Cllifford circuit models  to be discussed in Sec.~\ref{sec:circ}, and $(1+1)$-dimensional CFTs in the large central charge limit~\cite{Roberts:2014ifa}.
This was also used as a toy model in~\cite{abanin,MS}.
}. We will refer to this assumption as \slcg. See Fig.~\ref{fig:lc}.  From now on we will set $v_c =1$. Causality imposed by the light-cone structure will play a key role in the subsequent discussion. Throughout the paper, if not mentioned explicitly, we will always consider $q$ large and ignore subleading $1/q$ corrections. The qualitative picture does not change in more general situations, but the story becomes technically more complex, and will be treated elsewhere. 

The statement about light-cone growth does not say anything about the internal structure of an operator under time evolution. 
A basic question addressed in this paper is: can the operator develop a void in some region between its endpoints, and if so, what are the implications of this process?  
We say an operator has a void in some region $A$ (which can have disconnected components) if it is a superposition of operators the form 
$\tilde \sO \otimes \bid_A$, where $\bid_A$ denotes the identity operator in $A$ and $\tilde \sO$ is some operator  
which has nontrivial support in all disconnected parts of $\bar A$ (the complement of $A$).
Then the probability for a basis operator $\sO_\al$ to develop a void in region $A$ at time $t$ is 
\be \label{hn}
P_{\sO_\al}^{(A)} (t) = \sum_{\b \; \text{with void in $A$}}  |c_\al^\b (t) |^2 \ .
\ee

A main goal of this paper is to explore the role of void formation in the entanglement structure of a system. 
A good observable for this purpose is the evolution of the second Renyi entropy, which can be expressed in terms of operator growth ~\cite{abanin,MS,nahum1,frank}.  
Suppose at $t=0$, the system is described by a homogeneous pure product state, $\rho_0= \otimes_i \rho_i$,
where $i$ runs over all sites of the system and $\rho_i$ is a pure state which is the same for all sites. It is convenient to choose a basis so that 
\be \label{rhy}
  \rho_i= {1 \ov q} \le(\bid_i + Z_i + Z^2_i + \cdots + Z^{q-1}_i \ri)  
\ee
with $Z_i$ the clock matrix at site $i$, and thus 
\be  \label{heno}
\rho_0 = {1 \ov q^L} \sum_{\al \in I}  \sO_\al  
\ee
where $I$ denotes the set of operators which can be built from powers of $Z_i$'s. Note that the space $I$ is 
$q^L$-dimensional, in contrast to the $q^{2L}$-dimensional 
full space of operators. 

Under time-evolution, the reduced density matrix for some region $A$ is given by\footnote{For notational convenience we will take states of the system to evolve by $U^\da$.} 
\be \label{dens1}
\rho_A (t) = {\rm Tr}_{\bar A}  \rho (t) = 
 {1 \ov q^L} \sum_{\al \in I}{\rm Tr}_{\bar A} \sO_\al (t) =
{1 \ov q^{|A|}} \sum_{\al \in I} \sum_{\b \in A}  c_{\al}^\b (t) \sO_\b
\ee
where we have used~\eqref{evop}, and the fact that due to tracelessness of all nontrivial basis operators, only operators of the form $\sO_\b \otimes \bid_{\bar A}$ with $\sO_\b$ an operator in region $A$ (denoted by $\b \in A$)  contribute to 
${\rm Tr}_{\bar A} \sO_\al (t)$. $|A|$ denotes the size of region $A$. 
The second Renyi entropy for $A$ can then be written as 
\be \label{eej}
e^{-S_2^{(A)} (t)} = {\rm Tr}_{A} \rho_{A}^2 (t) =
 {1 \ov q^{|A|}}  \sum_{\al_1, \al_2 \in I} \sum_{\b \in A}  c_{\al_1}^\b (t) c_{\al_2}^{\b*} (t)  \ .
 \ee
 
We will now make a further simplification by ignoring the off-diagonal terms (i.e. terms with $\al_1 \neq \al_2$) 
in~\eqref{eej}. For chaotic systems, one expects the phases of $c_\al^\b$ to be random, so that the off-diagonal terms are suppressed by order $O(q^{-|A|})$ compared with diagonal terms~\cite{MS}. For integrable systems one cannot make this argument. 
Nevertheless, there are often situations where the off-diagonal terms vanish identically. This is the case for all the explicit examples we discuss in Sec.~\ref{sec:rnf} and Sec.~\ref{sec:circ}. We then find 
 \be \label{ejn}
e^{-S_2^{(A)} (t)} ={1 \ov q^{|A|}} N_A (t) %=  {1 \ov q^{|A|}} +  \hat P_{A} (t)
, \quad 
N_{A} (t) \equiv    \sum_{\al \in I} \sum_{\b \in A} | c_{\al}^\b (t)|^2  , \quad S_2^{(A)} = |A| \log q - \log N_A (t) \ .
  \ee
$N_A (t)$ has a simple physical interpretation: it is the expected number of operators in the set $I$  contained within region $A$
at time $t$. Note that $N_A \geq 1$,  as $O_{\alpha}= \mathbf{1}$ always contributes $1$ to the above sum. 
For our later discussion, it is convenient to introduce a function $N (A, B; t)$, defined as the expected number of initial operators in $I$ from some region $B$ that are contained in $A$ at time $t$, i.e. 
\be\label{nfu}
N (A, B; t) \equiv \sum_{\al \in I \cap B} \sum_{\b \in A} |c_\al^\b (t)|^2, 
 \ee
and a void formation function $G(A, B; t)$, defined as the expected number of initial operators in $I$ from some region $B$ that develop a void in $A$ at time $t$, i.e., 
\be\label{gfu}
G (A, B; t) \equiv \sum_{\al \in I \cap B} \sum_{\b \; \text{with void in $A$}} |c_\al^\b (t)|^2  = \sum_{\al \in I \cap B}  P_{\sO_\al}^{(A)} (t)
\ .
 \ee
$G(A, B; t)$ is closely related to $N (\bar A, B; t)$, but in $G(A, B; t)$ the final operators must be supported on all disconnected parts of $\bar A$. Again by definition,  $N (A, B; t), G(A, B;t)  \geq 1$, as we always have a contribution of 1 from the identity operator. 

Throughout the paper, we will denote the union of two regions $A_1 \cup A_2$  simply as $A_1 A_2$. 

\iffalse
Since the identity operator in $I$ always contribute $1$ to $N_A (t)$, it is convenient to separate it out and write 
\be 
N_A (t) = 1 + \hat N_A (t)
\ee
with $\hat N_A (t)$ denotes the contribution from non-identity operators in $I$. 
\fi

%\HL{Maybe also mention OTOC here} 

\subsection{{Upper bound on} average probability for void formation} 

To give some intuition for the motivation behind \eqref{pgd}, we first discuss the average probability for an operator to {become trivial in a given region}. 

Consider a Hilbert space $\sG$ of dimension $d$ and a subspace $\sG' \subset \sG$ of dimension $d'$, with 
$P$ the projector onto $\sG'$.  For a vector $\ket{\psi} \in \sG$, the probability that it transitions into a 
state in $\sG'$ under a unitary transformation is given by 
\be 
p = \Tr \le(P U \ket{\psi} \bra{\psi} U^\da \ri)  \ .
\ee
The average probability $\bar p$, obtained by averaging $\ket{\psi}$ over all unit vectors in $\sG$ with the unitarily invariant Haar measure, is then 
\be \label{nne}
\bar p = {1 \ov d} \Tr P = {d' \ov d} \ .
\ee
We take $\sG$ to be the Hilbert space of all operators with dimension $d = q^{2 L}$, and $\sG'$ to be the set of operators 
which are the identity in region $A$, which is a subspace of dimension $d'= q^{2 (L - |A|)}$. We thus conclude that 
the average probability for an operator to be trivial in region $A$ 
is 
\be\label{eba}
\bar p = q^{-2 |A|} = {1 \ov d_A^2} \ .
\ee
Apart from the assumption that $U$ is unitary,~\eqref{eba} is independent of $U$ and does not depend on the region $A$ other than through its size.
%\footnote{Note that the variance appears to be also independent of $U$ and $A$. (\HL{check})} %Thus it does not tell us anything about gap formation of  a generic operator.
Note that~\eqref{eba} gives the probability for an operator to be trivial in $A$, which is larger than the probability for developing a void in $A$, as it also includes final operators which are trivial in some disconnected parts of $\bar A$.

Note that if we do not average $\ket{\psi}$ but instead take $U$ to be a random unitary matrix with the Haar distribution, we find 
the same answer. This is the motivation for the name ``\rgd" \ for~\eqref{pgd}.

\subsection{Unitarity of entanglement growth for one interval} \label{sec:ent}

We now present a simple argument which shows that void formation plays a crucial role in 
ensuring that the entanglement growth of a system after a global quench is compatible with unitarity. The argument can be used to derive a constraint on void formation by requiring unitarity.  

\begin{figure}[!h]
\begin{center}
\includegraphics[width=16cm]{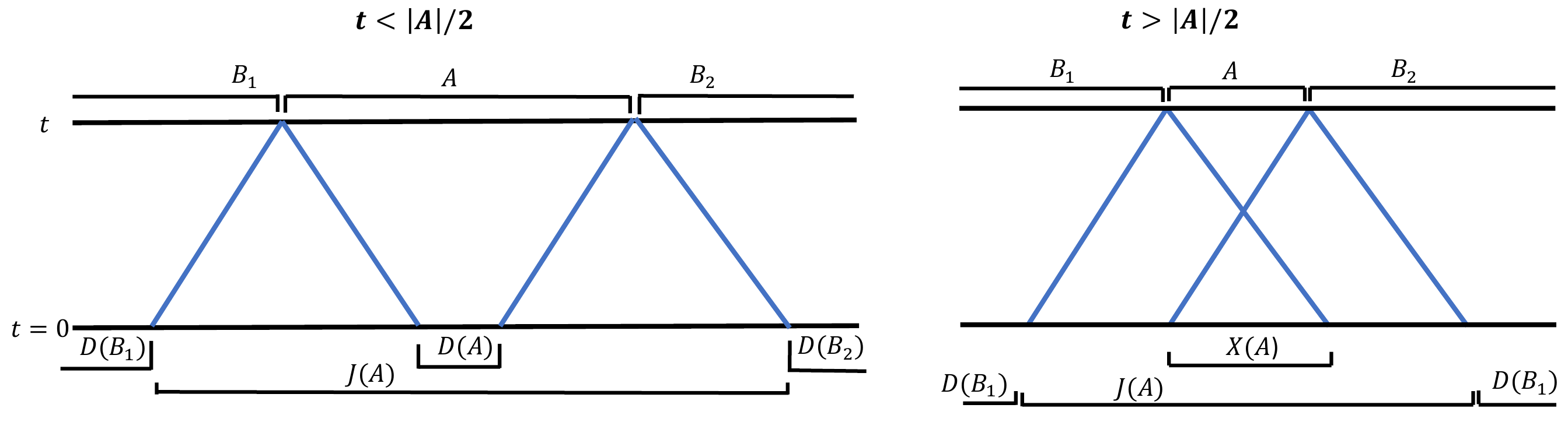} 
\caption{Left: $t < |A|/2$. Right: $t > |A|/2$.
$D(A)$ is the intersection of the past domain dependence of $A$ with the $t=0$ slice, which becomes empty for $t > |A|/2$.
$J(A)$ is the region at $t=0$ which is causally connected with $A$. At $t< |A|/2$, every operator in $J(A)$ {has zero probability of transitioning} to a final operator contained in $\bar{A} = B_1 B_2 $. At $t\geq |A|/2$, an operator in $J(A)$ can transition to an operator contained in $\bar A$ if it develops a void in $A$.}
\label{fig:regA}
\end{center}
\end{figure}

To find the second Renyi entropy~\eqref{ejn} for some finite interval $A$, we need to find the expected number of operators in $I$ which fall into $A$ as a function of time. 
The sharp light-cone growth of operators depicted in Fig.~\ref{fig:lc} makes the counting very simple~\cite{abanin,MS}. 
Let us denote the intersection of the past domain of dependence of region $A$ with the $t=0$ slice 
as $D(A)$. It then follows from the light-cone structure that all operators in $D(A)$  are contained within region $A$ at time $t$, while an operator with a nontrivial part outside $D(A)$ at $t=0$ will evolve into operators with nontrivial parts outside $A$. See Fig.~\ref{fig:regA}. Thus only the initial operators in $D(A)$ contribute to $N_A (t)$, 
and each of them contributes a probability $1$. $N_A (t)$ is then given by the number of basis operators in the set $I \cap D (A)$, which leads to
    \be \label{heb}
N_A (t) = \bca q^{|D(A)|} = q^{|A|-2t}  & t < {|A| \ov 2} \cr 1 & t > {|A| \ov 2} \eca \quad 
\Lra \quad S^{(A)}_2 (t)  = \bca 2 \seq t   &  t < {|A| \ov 2} \cr 
     \seq  |A| & t > {|A| \ov 2} 
     \eca , \quad \seq \equiv \log q \ .
\ee
Physically, under evolution, initial basis operators localized in $A$ move out at the lightcone speed in both directions. When an operator develops a nontrivial part outside $A$, it ceases to contribute to $N_A (t)$ and increases the entropy of $A$. At time $t$, the only operators remaining inside $A$ are those which are initially localized in $D(A)$. 
When $t > {|A| \ov 2}$, $D(A)$ is empty, with all nontrivial operators having evolved outside $A$.  
The only contribution to $N_A (t)$ is from the identity, and the entropy saturates. 
One can in fact show that with the \slcg , all the Renyi  and von Neumann entropies for a single interval $A$ are also given by~\eqref{heb}, and $\rho_A$ is unitarily equivalent to a reduced density matrix in which the region $A - D(A)$ is maximally entangled with $\bar A$ while $D(A)$ remains pure, as in the picture of the ``entanglement tsunami''~\cite{Liu:2013iza}.

This discussion provides a simple explanation for the linear growth of entanglement entropy and its saturation, and shows that such behavior has its physical origin in ballistic operator growth, regardless of whether the system is chaotic or integrable.\footnote{We will see some explicit examples for integrable systems in Sec.~\ref{sec:circ}.}
  But at this stage, there is an apparent violation of unitarity. Applying the above discussion to $S^{(\bar A)}_2 (t)$, we obtain the same behavior as~\eqref{heb} with $|A|$ replaced by $|\bar A| = L - |A| \to \infty$, which is inconsistent with  $S^{(A)}_2 (t)= S^{(\bar A)}_2 (t)$ for $t \geq |A|/2$. That is, instead of growing indefinitely, under unitary evolution $S^{(\bar A)}_2 (t)$ must also saturate at $\seq  |A|$ for $t > |A| /2$. 
%\footnote{\HL{Clearly if all operators grow in size under evolution, it can not be unitary. But it is ok to assume all the operators involved in a product pure state all 
%grow in size.}}   

The way out is as follows. Let us denote the region at $t=0$ which is in causal contact with $A$ as $J(A)$ (see Fig.~\ref{fig:regA}), and consider an initial operator of the form $\sO_{J(A)} \otimes \sO_{D(\bar A)}$ with $\sO_{J(A)}$ a nontrivial operator in $J( A)$.  
In the discussion above, such an operator was assumed to have no contribution to $N_{\bar A} (t)$, as naively its time evolution will be nontrivial in $A$.  But this is incorrect; such operators can contribute to $N_{\bar A} (t)$ if $\sO_{J( A)}$ develops a void in $A$. Moreover, due to causality, the behavior of $\sO_{J( A)}$ in region $A$ under time evolution--including the probability of developing a void--should be independent of $\sO_{D(\bar{A})}$. This means that $N_{\bar A} (t)$ can be written in a factorized form 
\be\label{nen}
N_{\bar A} (t) =  q^{|\bar A|- 2 t}  N (\bar A, J(A); t), % \quad 
%N (A, B; t) \equiv \sum_{\al \in I \cap B} \sum_{\b \in \bar A} |c_\al{^\b} (-t)|^2
\ee
where the factor $q^{|\bar A|- 2 t}$ is the number of basis operators in $I \cap D (\bar A)$, and 
the function $N$ was introduced in~\eqref{nfu}. When $A$ is a single interval, it is clear from causality that 
\be \label{hjk}
N (\bar A, J(A); t) = G (A, J(A); t)
\ee
where as defined in~\eqref{gfu}, $G (A, J(A); t)$ is defined as the expected number of initial basis operators in $J(A)$ that develop a void in $A$. 
For $S_2^{(A)} = S_2^{(\bar A)}$,  we need 
\be \label{heo}
G (A, J(A); t)  = \bca 1  & t < {|A| \ov 2} \cr q^{ 2t- |A|} & t > {|A| \ov 2}   \eca \ .
\ee
The second line of~\eqref{heo} has the simple interpretation that the average probability of the $q^{|J(A)|} = q^{|A|+2t}$ basis operators in $J(A)\cap I$ to develop a void in $A$ is $q^{-2|A|}$. Also note that $q^{2t -|A|} = q^{|X|}$, where the region $X$ is as shown in Fig.~\ref{fig:regA}. For an operator to develop a void in $A$, from causality it must be supported in region $X$. Note that while there are some similarities, equation~\eqref{heo} is  different in nature from~\eqref{eba}. Equation~\eqref{eba} is a purely kinematic statement, while~\eqref{heo} contains the dynamical input of the \slcg\ and refers to an average over a more restricted set of initial operators. 

Now consider the following ``Renyi  mutual information'' between regions $B_1$ and $B_2$ in Fig.~\ref{fig:regA}: 
\be
I_2 (B_1, B_2) \equiv S_2^{(B_1)} + S_2^{(B_1)}  - S_2^{(\bar A)} , \quad \bar A = B_1  B_2 \ . 
%= \bca 0 & t < {|A| \ov 2} \cr
   %                                    \seq (2 t - |A|) & t > {|A| \ov 2} 
      %                                 \eca \ .
\ee
From~\eqref{nen} and~\eqref{hjk}, we find that this quantity is fully controlled by the expected number of operators developing a void
\be \label{kej}
I_2 (B_1, B_2) = \log G (A, J(A); t) = \bca 0 & t < {|A| \ov 2} \cr
                                       \seq (2 t - |A|) & t > {|A| \ov 2} 
                                       \eca \ .
\ee
This result is intuitively appealing: when an operator develops a void in an interval $A$, it leads to mutual information between regions separated by $A$. 
 
We stress that~\eqref{heo}, and accordingly~\eqref{kej}, are constrained by unitarity and should apply to any system, integrable or chaotic, which has \slcg\
for the initial set of operators. 
We will see that~\eqref{heo} is indeed satisfied in various exactly solvable unitary circuit models in Sec.~\ref{sec:rnf} and Sec.~\ref{sec:circ}.

\subsection{Mutual information and multi-partite entanglement} \label{sec:mutual}

Here we explore further implications of void formation, as well as general constraints on this process, 
by examining the Renyi entropy for two and more disjoint intervals. Extending~\eqref{kej}, we first show that the mutual information between two disjoint finite intervals is determined by the void formation function $G (B, \tilde B; t)$  for some appropriate regions $B$ and $\tilde B$.  In a certain region of the parameter 
space, the corresponding mutual information is universal, determined by~\eqref{heo} from unitarity.
% but in general the gap formation function and the mutual information between finite intervals are model dependent. 
We also derive new constraints on void formation from unitarity of entanglement growth for two intervals. For more disjoint intervals, we see that 
void formation in operator growth leads to multi-partite entanglement, and void formation functions provide new measures for 
characterizing multi-partite entanglement.

\begin{figure}[!h]
\begin{center}
\includegraphics[width=14cm]{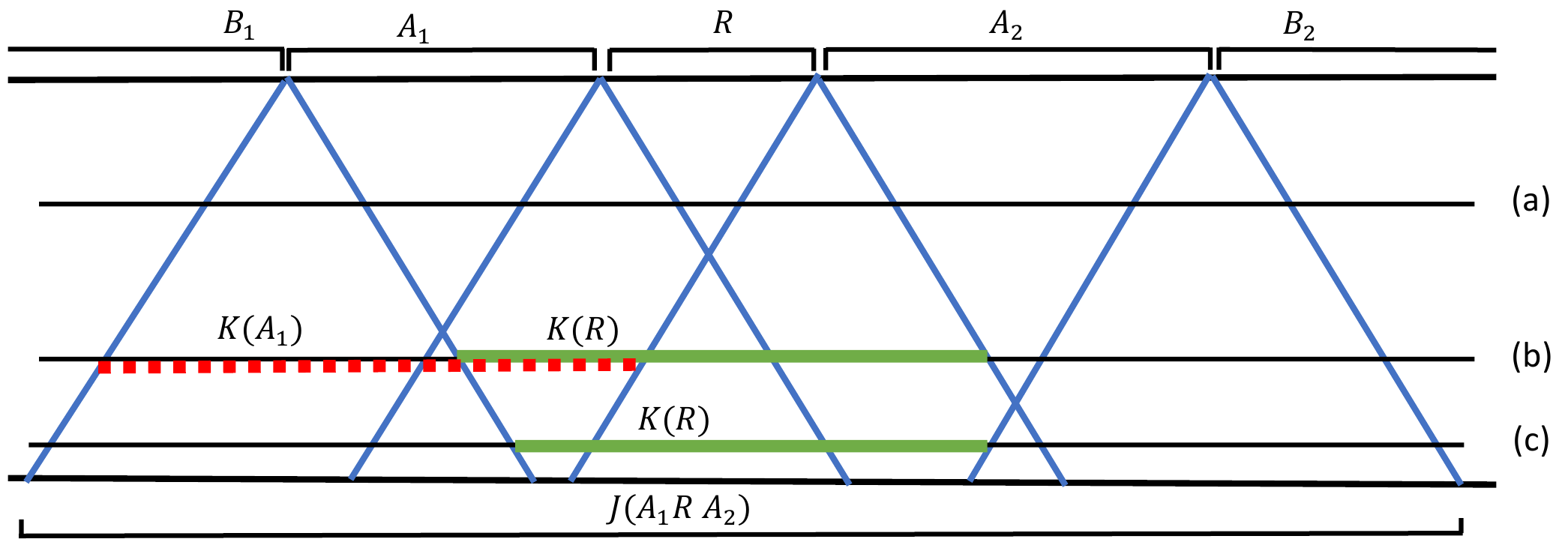} 
\caption{Different situations for two intervals. Different horizontal lines correspond to $t=0$ slices for (a) $t < |R|/2$; (b) $t > |l_1|/2$; (c) $t > |l_2|/2$. Here we take $|R|< |A_1|$ and thus $l_1 = |A_1|$ and $l_2 = |A_2|$.}
\label{fig:two}
\end{center}
\end{figure} 

Consider $S_2$ for a region $A= A_1  A_2$ separated by an interval $R$. Without loss of generality, we can take $|A_1| \leq  |A_2|$. For $t < |R|/2$, from causality, $N_A (t)$ of~\eqref{ejn} factorizes into a product of the functions for $A_1$ and $A_2$ 
\be\label{enne}
N_A (t) = N_{A_1} (t) N_{A_2} (t) , \quad S_2^{(A)} (t) =  S_2^{(A_1)} (t) +  S_2^{(A_2)} (t) , \quad t < {|R| \ov 2} 
\ee
where $N_{A_1} (t)$ denotes the contribution from initial operators in the region $D(A_1)$ and is given by~\eqref{heb}. 
For $t > {|R| \ov 2}$, initial
operators in the region $K (R) = J(R)  \cap D(A_1 R  A_2) $ (see Fig.~\ref{fig:two}) can now potentially form a void in region $R$ and thus contribute to $N_A (t)$ 
\be\label{myu0}
N_A (t) =  N_{A_1} (t) N_{A_2} (t) G(R, K (R); t)   , %\qquad %L = \tilde R \, \cap D(A_1 \cup R \cup A_2)
\ee
which gives 
\be \label{myu}
I_{2} (A_1, A_2; t) =  \log G (R, K (R); t)  \ .
\ee

For further discussion, it is convenient to introduce the following notation: 
\be 
l_1 \equiv {\rm max} (|A_1|, |R|), \quad l_2 \equiv {\rm max} (|A_2|, |R|), \quad
K(A_1) \equiv J(A_1) \cap D (B_1 A_1 R) % \quad K(A_2) = J(A_2) \cap D ( R A_2B_2) 
\ .
\ee

Note that when $t< |R|/2$, we have 
the factorized form~\eqref{enne} in any theory with sharp light-cone growth. For $|A_1|/2 > t > |R|/2$ (which can happen for $|A_1| > |R|$), we have 
$K (R)= J( R)$, and from~\eqref{heo}
\be 
I_{2} (A_1, A_2) = \seq (2t -|R|)  \ .
\ee
 For $t > {|A| + |R| \ov 2}$, $K(R)$ becomes an empty set, and thus $G(R, K(R); t)  =1$. So for $t< l_1/2$ and for  $t > {|A| + |R| \ov 2}$, $G (R, K (R); t)$ and $I_{2} (A_1, A_2)$ have a universal form in all systems with sharp light-cone growth\footnote{In the discussion of~\cite{Asplund:2015eha} of Renyi entropies for two-dimensional conformal field theories (CFTs), these are indeed the regimes which are universal for all CFTs.}. For ${|A| + |R| \ov 2} > t > l_1/2$,
$K (R) \subset J(R)$,  the behavior of $G (R, K (R); t)$ becomes system-dependent, and so does $I_{2} (A_1, A_2)$.  
Now consider the entropy for the region $\bar A = B_1 R B_2$, for which  
\be \label{oyh}
N_{\bar A} (t) =   N_{B_1} (t) N_{B_2} (t) N (\bar A, J(A_1 R A_2);t)  \ .
%N (\bar A, J(A_1RA_2); t) 
\ee
Similar to the discussion immediately above, for $t \leq l_1/2$, the constraint~\eqref{heo} is enough to ensure $S_2^{(A)} =   S_2^{(\bar A)}$, but for $t > l_1/2$ new constraints arise. 
We find (see also Fig.~\ref{fig:two}) 
\bea \label{ot1}
& {l_1 \ov 2} < t \leq {l_2 \ov 2}: \quad & G(R, K(R); t) = q^{|A_1| - |R|} G (A_1, K(A_1) ;t)  \\
& t > {l_2 \ov 2}: \quad & G(R, K(R); t) = q^{|A| - |R|-2t} N(\bar A, J(A_1 R A_2); t),
\label{ot2} 
\eea
For $t > {|A| + |R| \ov 2}$,  with $G(R, K(R); t)  =1$, we must have
\be\label{ot3}
N(\bar A, J(A_1 R A_2); t) = q^{ 2t + |R|- |A|}, \quad t > {|A| + |R| \ov 2} \ . 
\ee

%\HL{Since $A_2$ can be made arbitrarily large for fixed $A_1$ and $R$, \eqref{ot1} should be viewed as a relation between the void formation functions of two adjacent intervals that is valid for arbitrary $t> l_1/2$.}

\begin{figure}[!h] 
\includegraphics[width=12cm]{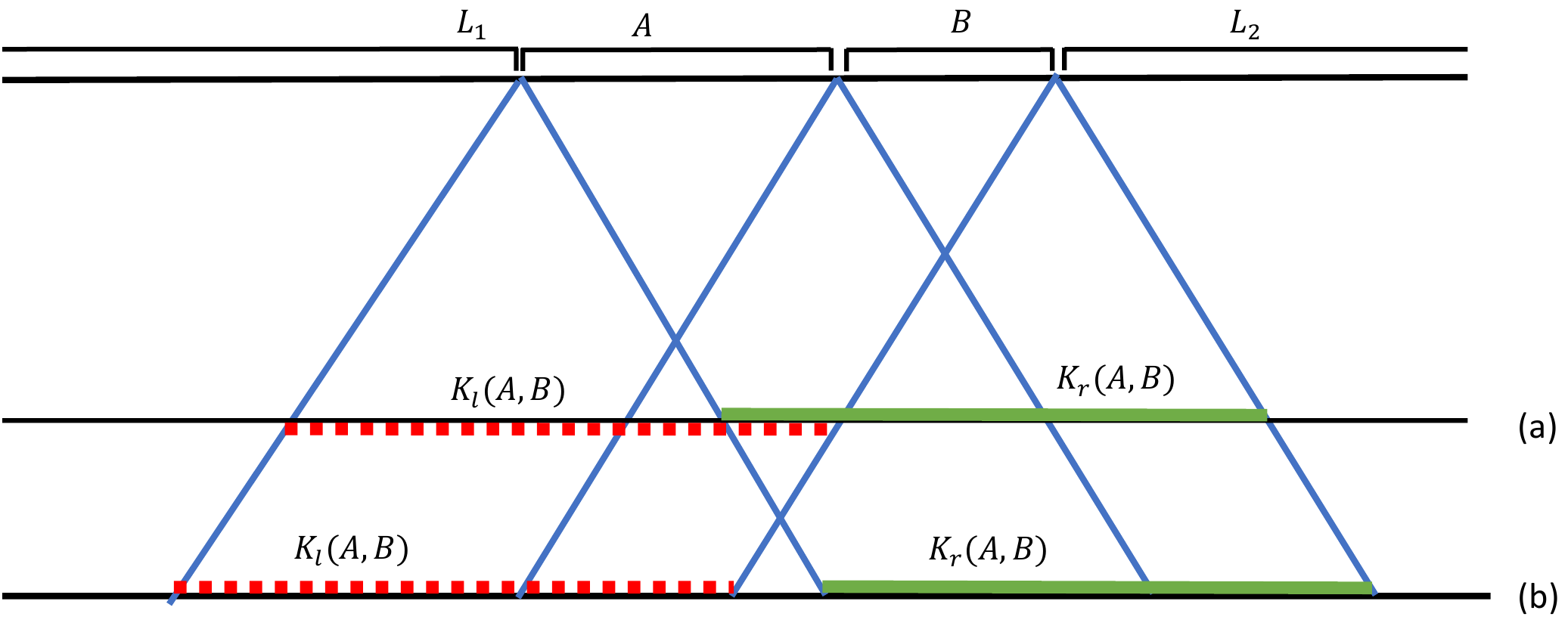}
\caption{Regions $K_l(A, B)$ and $K_r(A, B)$ of \eqref{AB_condition} at a time $t<(|A|+|B|)/2$ in (a), and $t>(|A|+|B|)/2$ in (b). The condition \eqref{AB_condition} holds at all times $t> \text{max}(|A|/2, |B|/2)$.}
\label{fig:AB_condition}
\end{figure}

{If we take $A_2$ to be the entire semi-infinite region to the right of $R$, 
then both regions $K(A_1)$ and $K(R)$ appearing in \eqref{ot1} and \eqref{ot2} depend only on $A_1$ and $R$, and $l_2/2\rightarrow \infty$, so \eqref{ot1} holds for all $t> l_1/2$. We can thus deduce from \eqref{ot1} a general constraint on the void formation functions for any two adjacent finite intervals $A$ and $B$ in an infinite system, 
\be 
q^{|A|} G(A, K_l(A, B); t) = q^{|B|} G(B, K_r(A, B); t), ~~~~ t> \text{max}(|A|/2, |B|/2)  \label{AB_condition}
\ee
where 
\be 
K_l(A, B) = J(A)\cap D(L_1 A B), ~~~~ K_r(A, B) = J(B)\cap D(A B L_2)
\ee 
where $L_1, L_2$ are the semi-infinite regions to the left of $A$ and to the right of $B$ respectively. The different regions appearing in \eqref{AB_condition} are shown in Fig.~\ref{fig:AB_condition}.}

The above discussion can be generalized to 
express the second Renyi entropy for any number of intervals in terms of appropriate void formation functions. 
On the one hand, by requiring $S_2^{(A)}(t) = S_2^{(\bar{A})}(t)$ for $A$ consisting of an arbitrary number of intervals, one can obtain further constraints on the void formation functions.
On the other hand, with the full knowledge of the void formation functions, one should be able to deduce the expression for $S_2^{(A)}$ for any $A$. We will see an explicit example of this in Sec.~\ref{sec:max}. 

\begin{figure}[!h]
\begin{center}
\includegraphics[width=12cm]{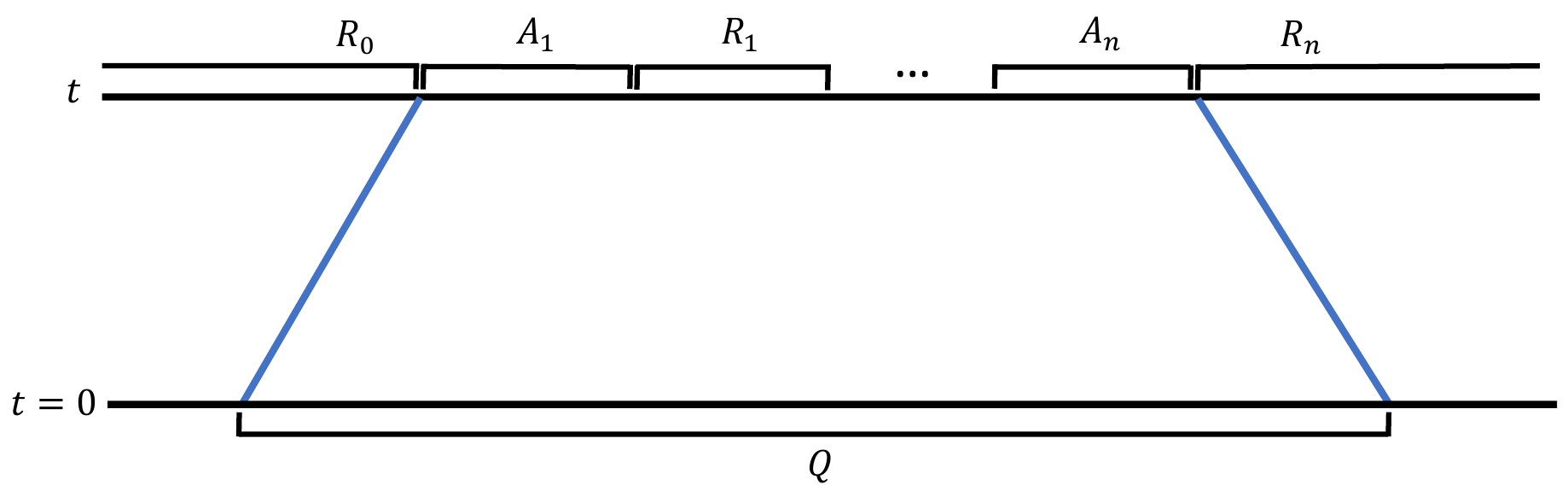} 
\caption{When initial operators in $Q$ develop a void in $A = A_1 \cdots A_n$, this leads to multiple-partite entanglement among regions $R_0, R_1, \cdots R_n$. Here $Q$ is {$J(A_1 R_1 \cdots R_{n-1} A_n)$.}}
\label{fig:n-int}
\end{center}
\end{figure}

Now consider a region consisting of $n$ disjoint intervals $A= A_1 \cdots A_n$ separated by intervals $R_1, \cdots R_{n-1}$, as in Fig.~\ref{fig:n-int}. Then the void formation function $G(A, Q; t)$ gives a contribution to the entropy of the region $\bar A = R_0 R_1 \cdots R_{n}$, but does not contribute to the entropy of any region consisting of a proper subset of $\{R_0, \cdots, R_n\}$. 
Thus void formation in $A$ leads to multi-partite entanglement among all disconnected regions in $\overline{A}$, which can be captured by the quantity $G(A, Q; t)$.

\section{Random void distribution and entanglement growth} \label{sec:rnf}

In this section, we consider the probability distribution of void formation for a generic operator 
in the random unitary circuit model in the large $q$ limit. We show that it is given by the random void distribution~\eqref{pgd}.
We then show that {by assuming} the  \rgd\ {for all initial operators, we can correctly obtain}   the full expression for $S_2^{(A)}$ in the random circuit model for $A$ consisting of an arbitrary number of disjoint intervals. Surprisingly, the resulting expression is found to coincide with the von Neumann entropy  for holographic systems. 
%\HL{On the one hand, this strongly suggests that the \rgd\ underlies the entanglement growth of holographic systems.}
%\HL{On the other hand, from the earlier discussion that the holographic expression maximizes the entanglement entropy~\cite{hong_mark},
%it implies that together with \slcg, the \rgd\ leads to maximal entanglement growth.} 
In the next section, we will contrast the 
\rgd\ with the void formation properties of two Clifford circuit models~(which may be seen as non-chaotic systems).

\subsection{Random unitary circuits} 

\begin{figure}[!h]
\begin{center}
\includegraphics[width=14cm]{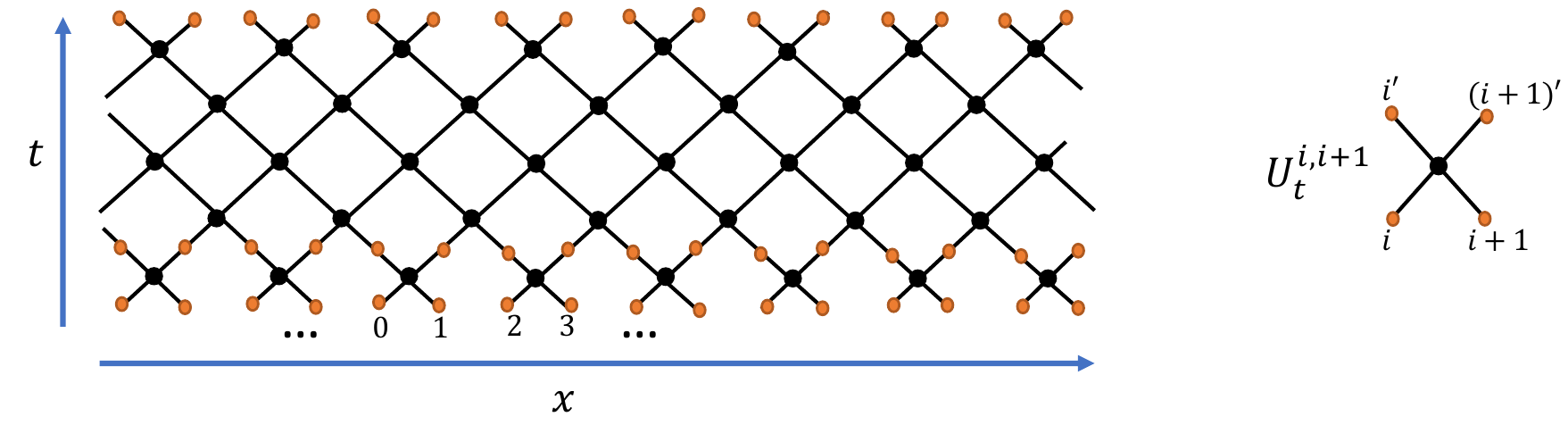} 
\caption{Unitary circuit for the time-evolution operator $U(t)$~\eqref{11}--\eqref{12} for $t=5$. The black circles represent % four-index tensors that act as 
unitary matrices $U^{i, i+1}_t$ acting on two adjacent sites $i$ and $i+1$. %Lines between the tensors represent index contractions. 
In the random circuit model, $U^{i, i+1}_t$ for each $i$ and $t$ is independently drawn from the Haar ensemble of $q^2\times q^2$ unitary matrices. Here in the first step of the time-evolution, we couple sites 0 and 1, 2 and 3, and so on, and in the second step we couple -1 and 0, 1 and 2, and so on. We explicitly show lattice sites of the system at $t=0$ and $t=1$ with the orange circles.}
\label{fig:circuit_structure}
\end{center}
\end{figure}

We first describe briefly the setup of the random unitary circuit discussed in~\cite{nahum1,frank,nahum2}, and 
its main properties. Consider a time-evolution of the system as in Fig.~\ref{fig:circuit_structure}, where the evolution operator 
$U(t)$ can be written as  
\bega \label{11}
U (t) = U_{t} U_{t-1} \cdots U_0, \\U_0 = \cdots \otimes U_0^{0,1} \otimes U_0^{2,3} \otimes \cdots , \quad 
U_1 = \cdots \otimes U_1^{-1,0} \otimes U_1^{1,2} \otimes \cdots \ .
\label{12}
\end{gather} 
Here we take discrete time steps, with $U_n$ corresponding to the evolution operator at the $n$-th step. 
$U^{i,i+1}_n$ is a $q^2 \times q^2$ unitary matrix that acts on two neighboring sites $i$ and $i+1$ 
at $n$-th step, and each such matrix is averaged over the Haar measure independently. 
We will consider the limit of large $q$. 

The random unitary circuit  can be seen as a discretized model for 
chaotic quantum systems. It is manifestly unitary and local, but replaces the 
local interactions between neighboring spins in a realistic Hamiltonian system by random ones. 
It provides a powerful playground for studying chaotic systems, as many observables 
such as entanglement entropy, OTOCs, and operator spreading coefficients are analytically calculable, and the resulting behavior 
has been found to be consistent with numerical results of realistic chaotic spin-chain systems~\cite{nahum1,frank, swingle, swingle2, levi}.

Here are some features of the random circuit model in the large $q$ limit which are relevant for our discussion~\cite{nahum1,frank,nahum2}: 

\ben 

\item For a basis operator $\sO_\al$ (introduced above~\eqref{hne}) with
 right and left endpoints given by $x_r$ and $x_l$ respectively, one finds
\begin{equation}
\begin{gathered}
    \sum_{\substack{\b \text{ with left endpoint} \\ x_l-t,  \text{ right endpoint } x_r+t}} \overline{\le| c_{\al}^\b(t)\ri|^2} = 1 + O(1/q)
\label{op_growth_1}
\end{gathered}
\end{equation}
which implies that the end points of any operator $\sO$ move under time evolution in opposite directions with light cone speed 
$1$ as in Fig.~\ref{fig:lc}. Here and below, an overline denotes an average over the random unitaries. 

\item The calculation of the second Renyi entropy $S_2^{(A)}$ for a region $A$ can be reduced to computing the partition function of a classical Ising model on a triangular lattice. For $A$ consisting of a single interval, it has 
the form~\eqref{heb}, consistent with the general argument presented earlier. 

%In Appendix~\ref{app:renyi} we generalize the calculation for $A$ consisting of arbitrary disjoint intervals (see more on this below). 

\item In the large $q$ limit, 
\be 
e^{-\overline{S_2^{(A)}}} =  \overline{e^{-S_2^{(A)}}} = {1 \ov q^{|A|}} \overline{N_A (t)} 
\ee
where $N_A (t)$ was defined in~\eqref{ejn}. 
Here the off-diagonal terms in~\eqref{eej} automatically vanish due to random averages. 
Furthermore, one has 
\be\label{uno}
\overline{S_2^{(A)}} = \overline{S_n^{(A)}} = \overline{S^{(A)}} , \quad \forall n > 2  , \quad n \in \NN
\ee
where $S^{(A)}$ denotes the von Neumann entropy. So our discussion below about $S_2$ can also be understood 
as being relevant for the von Neumann entropy. 

\een

\subsection{Random void distribution}  \label{sec:rgd2}

We now look at the probability distribution for a generic operator $\sO$ %\textcolor{red}{without a void} 
to develop a void in some designated region $A$ in 
the random circuit model at large $q$. 

We can expand $\sO$ in terms of basis operators as 
\be \label{nmb}
\sO = \sum_\al a_\al \sO_\al , \qquad \sO (t) = \sum_\al \sum_\b a_\al c_\al^\b (t)  \sO_\b, \quad \sum_\al |a_\al|^2 =1. 
\ee
Then under time evolution, the probability for $\sO$ to have a void in $A$ is 
\be \label{hn1}
\overline{P_\sO^{(A)} (t) } = \sum_{\b \, \text{with void in $A$}}  \sum_{\al_1, \al_2} a_{\al_1} a_{\al_2}^* \overline{c_{\al_1}^{\b} (t) c_{\al_2}^{*\b} (t)} = \sum_\al |a_\al|^2  \overline{P_{\sO_\al}^{(A)} (t)}
%= \sum_{\b \, \text{with void in $A$}} \sum_\al |a_\al|^2  |c_{\al}^{\b} (t)|^2 
\ee
where $P_{\sO_\al}^{(A)}$ was introduced in~\eqref{hn}, and in the second equality the off-diagonal terms drop out due to the random average.   In~\eqref{hn1} by ``$\b$ with void in $A$'' 
we mean $\sO_\b$ should be trivial in $A$ and have support in each disconnected part of $\bar A$.
For instance, in the case of $A = A_1A_2 \cdots  A_n$ in Fig.~\ref{fig:n-int}, $\sO_\b$ should be the identity in $A$ while being nontrivial in each of $R_i$'s. From now on, for notational simplicity we will suppress the explicit overline for averages.  

\begin{figure}[!h]
\begin{center}
\includegraphics[width=12cm]{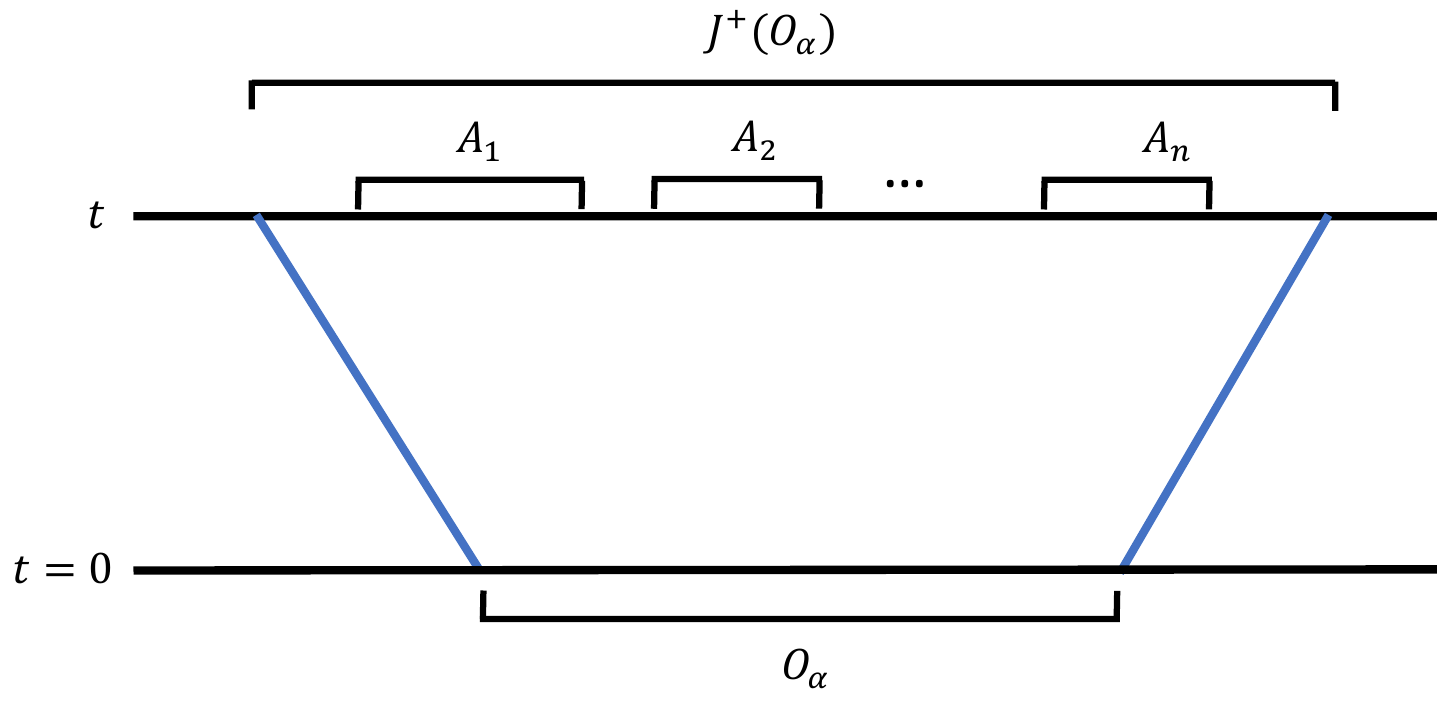} 
\caption{Regions $J^{+}(\sO_{\alpha})$ and $A_1, A_2, ..., A_n$. In the random void distribution, the probability of forming a void in $A \equiv A_1A_2...A_n$ depends only on the lengths $|A_1|, ..., |A_n|$.}
\label{fig:random_gap}
\end{center}
\end{figure}

$P_{\sO_\al}^{(A)} (t)$ can be expressed as the partition function of a classical Ising model on a triangular lattice with boundary conditions specified by $\sO_\al$ and $A$. We present the details of the calculation in Appendix~\ref{app:rgd}. For any operator $\sO_\al$ which does not  have an initial void, the final result 
is the random void distribution (RVD) already mentioned  in the Introduction section 
\be \label{rgd1}
P_{\sO_\al}^{(A)} (t)  = \bca e^{-2 \seq |A|} = {1 \ov d_A^2} &  t \geq {\rm max} \le({|A_i| \ov 2} , \forall i \ri),  A \in J^+ (\sO_\al) \cr
  0 & \text{otherwise} 
  \eca
\ee
where we have taken $A = A_1A_2 \cdots  A_n$ with $|A| = |A_1| + \cdots + |A_n|$, 
$J^+ (\sO_\al)$ denotes the region at time $t$ which is causally connected to $\sO_\al$, and $\seq = \log q$.  See Fig.~\ref{fig:random_gap}.
That to have a nonzero result we must have $ A \in J^+ (\sO_\al) $ follows simply from causality. 
If any of the $A_i$ are not in $J^+ (\sO_\al) $, clearly $\sO_\al$ cannot evolve into an operator which is nontrivial 
in all disconnected parts of $\bar A$.

Operators with initial voids can in general evolve to final operators that have support in all disconnected parts of $\bar{A}$ by two distinct kinds of processes. We can have processes where at some intermediate stage, the operator evolves to an operator without a void (if the initial voids are denoted by $G_1, ..., G_m$, such processes are allowed by causality only at $t>|G_i|/2$ for all $i$), which then splits to form a void in $A$. Another kind of process is one where different disconnected parts of the initial operator evolve to different disconnected parts of $\overline{A}$ without interacting with one another. Depending on the sizes and locations of the $G_i$, either of these types of processes can give the dominant contribution to  $P_{\sO_\al}^{(A)} (t)$ in the large $q$ limit. When the leading contribution is given by the former process, which can physically be seen as one of genuine ``void formation," we still have \eqref{rgd1}. When one of the latter types of processes is dominant, $P_{\sO_\al}^{(A)} (t)$ is enhanced compared to~\eqref{rgd1}.  Nevertheless, for a generic operator~\eqref{nmb}, the part of the operator corresponding to $\sO_\al$ with an initial void is suppressed due to a much smaller phase space. One can show that the enhancement from disconnected processes never overcomes the suppression. See Appendix~\ref{app:rgd}. 
Given that~\eqref{rgd1} is independent of $\sO_\al$ other than through the causality constraint $A \in J^+ (\sO_\al)$, a generic operator $\sO$ also satisfies~\eqref{rgd1}.

Note that if we assumed that for a given initial operator, the probability of having any of the $q^2$ basis operators at any site between the endpoints of the operator at time $t$ is the same, then the probability $q^{-2 |A|}$ would immediately follow:
having a void in $A$ corresponds to fixing the operators at $|A|$ sites in the final operator to be one of $q^2$ options, while allowing the remaining operators to take any value. A similar ergodicity assumption was used in the operator growth model introduced in~\cite{abanin}. Equation~\eqref{rgd1} also applies to the random circuit model at a finite $q$ 
in the regime that region $A$ {and $t$ are} large (so that $q^{|A|}, q^{t}$ are large), as we will discuss elsewhere. 

From~\eqref{rgd1} one can obtain the void formation function~\eqref{gfu} for any regions $A$ and $B$. 
In the simplest case with $A$ consisting of a single interval lying within the future light cone of a region $B$, 
we have 
\be \label{unen}
G_{\rm RVD}(A, B; t) = \bca 1 &  |B| \leq 2 |A| \text{ or } \; t \leq {|A| \ov 2} \cr
                     e^{\seq (|B| - 2 |A|)}  &  |B| > 2 |A|  \text{ and } \; t > {|A| \ov 2}
                     \eca  \ .
\ee
We can check that the above expression satisfies the unitarity constraint ~\eqref{heo} by taking $B = J(A)$, so that $|B| = |A| + 2t$.  {We can similarly check that it satisfies the constraints \eqref{ot1} and \eqref{ot2}.}

For the Renyi entropy of two intervals~\eqref{myu0}--\eqref{myu},  we need $G (R, K(R); t) $ 
with $K (R) = J(R)  \cap D(A_1 R  A_2)$, whose behavior depends on the relative sizes of $|A_1|$, $|A_2|$ and $R$. 
For example, for $|R|< |A_1|$ we find that 
\be 
G (R, K(R); t) = \bca 1  & t \leq {|R| \ov 2} \; \text{or} \;  t > {|A_1 + |A_2| - |R| \ov 2}    \cr
              q^{2t- |R|}  & {|A_1| \ov 2} \geq t \geq {|R| \ov 2}  \cr
              q^{|A_1|- |R|}   & {|A_2| \ov 2} \geq t \geq {|A_1| \ov 2} \cr
              q^{|A_1| + |A_2| -2t -|R|} & {|A_1 + |A_2| - |R| \ov 2} \geq t \geq {|A_2| \ov 2} 
                            \eca , 
\ee
which upon using~\eqref{myu0} leads to 
\be \label{unj0}
S_2^{(A)} (t) =\seq  \bca  4t  & t \leq {|R| \ov 2} \cr 
                  |R| + 2t &  {|A_1 + |A_2| - |R| \ov 2}\geq t \geq {|R| \ov 2}  \cr
            %     2t + |A_1| - |A_1| + |R| = |R| + 2t & \geq t \geq {|A_1| \ov 2} \cr
              %   |A_1| + |A_2| + 2t + |R| - |A_1| - |A_2| = |R| + 2t & {|A_1 + |A_2| - |R| \ov 2} \geq t \geq {|A_2| \ov 2} \cr
                  |A_1| + |A_2| &  t > {|A_1 + |A_2| - |R| \ov 2}  
                  \eca  \ . 
\ee
For $|A_1|< |R|$ we find that 
\be \label{unj}
G (R, K(R); t) = 1, \quad \Lra \quad S_2^{(A)} (t) = S_2^{(A_1)} (t) + S_2^{(A_2)} (t)   \ .
\ee
The mutual information between $A_1$ and $A_2$ is thus nonzero only for the case $|R| < |A_1|$. 
% it can explicitly checked to be the case for two intervals by using~\eqref{rgd1} for~\eqref{oyh}. 
One can also obtain $S_2^{(A_1 A_2)}$ using the partition function method of Appendix~\ref{app:renyi}, 
and one finds agreement with~\eqref{unj0}--\eqref{unj}.

An alert reader may notice that the expressions~\eqref{unj0}--\eqref{unj} coincide with the expressions for the evolution of entanglement entropy after a global quench  
in holographic systems~\cite{Balasubramanian:2011at,Leichenauer:2015xra}. We will see below that this is not an accident; the results agree for any number of intervals.   

\begin{figure}[!h]
\centering 
\includegraphics[width=10cm]{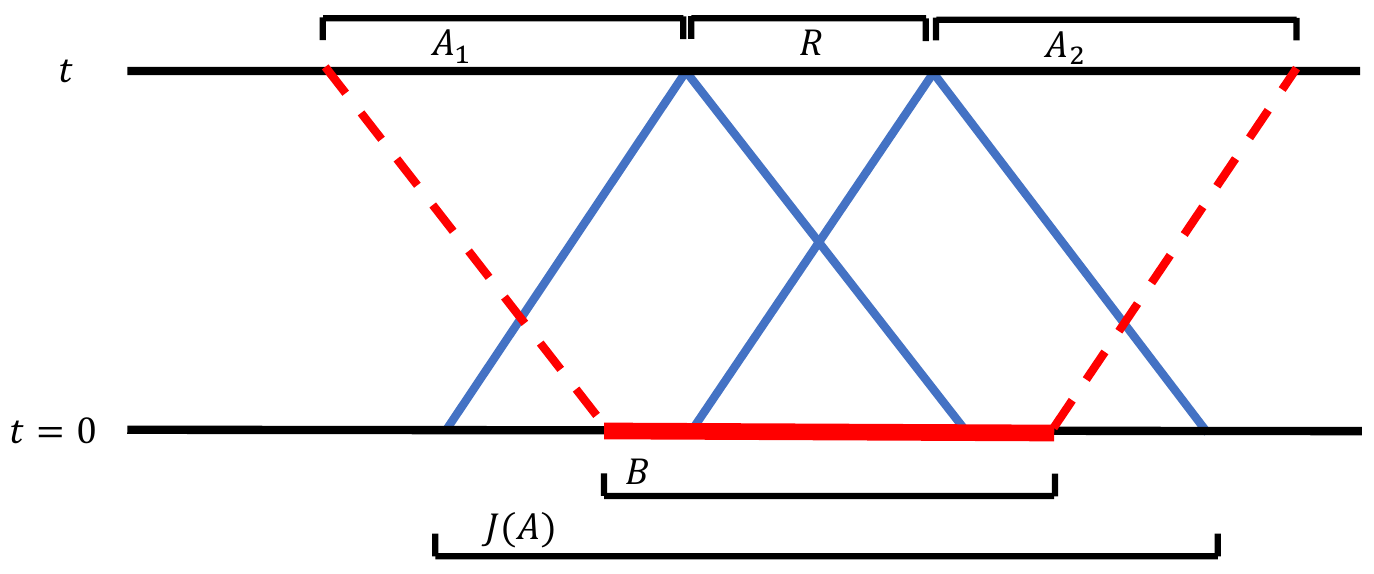}
\caption{For any $B$ in $J(R)$ such that $R$ is in the light-cone of $B$, we can construct $A_1, A_2$ such that $K(R) \equiv D(A_1 R A_2)\cap J(R) =B $. Then if $G(R, B, t)$ does not agree with~\eqref{unen}, then using \eqref{myu0}, we see that the resulting $S_2^{(A_1 A_2)} (t)$ 
will deviate from~\eqref{unj0}--\eqref{unj}, and thus from the holographic expression.}
\label{fig:general_A1A2}
\end{figure}

To conclude this subsection, we note that if there exists some $B$ for which $G(A, B; t)$ does not have the value 
in~\eqref{unen} for some $A$ in the light-cone of $B$, then we can always construct intervals $A_1$ and $A_2$ for which $S_2^{(A_1 A_2)}(t)$ deviates from~\eqref{unj0}--\eqref{unj}. See Fig.~\ref{fig:general_A1A2}. Thus   $S_2^{(A_1 A_2)}(t)$ agrees with the holographic result  for all $t$ and all $A_1, A_2$ if and only if \eqref{unen} is satisfied. 

 \subsection{Random void distribution and maximal entanglement growth} \label{sec:max}

\iffalse
\textcolor{red}{[}
As discussed at the end of Sec.~\ref{sec:mutual}, with the  knowledge of void distributions, one should be able to
deduce the entanglement entropy for any number of intervals. We now show that this is indeed the case for the random circuit  model, and also relate the result of the random circuit  model
to that of holographic systems. We leave most details to the appendices, and only present the main results here. \textcolor{red}{] Maybe we can remove this introductory paragraph, as it is not quite true that we are using the exact void probability distribution in random unitary circuits here, and we already outlined what we are doing on page 18?}
\fi

Consider a region $A = A_1  \cdots  A_n$ consisting of $n$ intervals $A_1 = [l_1, r_1], \cdots, A_n=[l_n, r_n]$, separated by intervals $R_1, ..., R_{n-1}$, as in Fig.~\ref{fig:n-int}.  The entanglement entropy $S_2^{(A)}$ can be calculated using the partition function method, and is discussed in detail in Appendix~\ref{app:renyi}. The result can be written as 
\begin{equation} 
S_2^{(A)} (t) = \seq ~ \min_{\{\gamma\}(t)} \bigg[  n_{\gamma} t + \sum_{\{l_i, r_j\}\in \gamma} | l_i - r_j| \bigg]
\label{hol2}
\end{equation} 
where the set $\{\gamma\}(t)$ and $n_\ga$ are defined as follows. {Call any pair of left and right endpoints $\{l_i, r_j\}$  from the 
set of all endpoints $\{l_1, r_1, ..., l_n, r_n\}$ a connectable pair at time $t$ if $t> | l_i-r_j|/2$. We can then group together the 
elements of $\{l_1, r_1, ..., l_n, r_n\}$ into different configurations such that every element is either unconnected to any other point, or part of one connected pair. Only connectable pairs can be connected. $\{\gamma\}(t)$ is the set of all such configurations $\gamma$ at time $t$, and $n_{\gamma}$ is the number of unconnected points in the 
configuration $\gamma$. Each connected pair $\{l_i, r_j\}$ in a configuration $\gamma$ contributes $|l_i-r_j|$, while each unconnected point contributes $t$, so that we get \eqref{hol2}.}   

Extending our discussion in section \ref{sec:rgd2} for two intervals, one can show that the same expression~\eqref{hol2} follows by using only the following elements: (i) \slcg; (ii) the \rgd~\eqref{rgd1} for all initial operators;  (iii) large $q$. 
The derivation is given in Appendix~\ref{app:alt}. 
%\textcolor{red}{[}This confirms  the expectation that void formation distributions can be used to ``determine'' the full structure 
%of $S_2^{(A)}$ for any $A$. \textcolor{red}{]}

The expression~\eqref{hol2} can also be shown to be equivalent to the expression of  entanglement entropy (in a scaling regime) for holographic systems after a quench. Holographic systems are a certain class of strongly coupled $(1+1)$-dimensional conformal field theories (CFT) which have a gravity dual. In the large central charge limit, their entanglement entropy can be calculated using classical gravity. More explicitly, in the regime 
with $t, |A_1|, \cdots ,|A_n|, |R_1|, \cdots, |R_{n-1}|$ large while $|A_i|/t, |R_i|/t$ are fixed,  
the holographic entanglement entropy after a global quench has the following form~(see e.g.~\cite{hong_mark})
\begin{equation} 
S_{\rm hol} (t) = \min\limits_{\sigma} ~ \bigg[\sum_{i=1}^n S_{\text{interval}}(t, |l_i- r_{\sigma(i)}|) \bigg]
\label{hol1}
\end{equation} 
where 
\begin{equation} 
S_{\text{interval}}(t, R) = \begin{cases} 2 \seq t & t < {R \ov 2} \\
\seq R &  t \geq {R \ov 2} 
\end{cases}
\label{holint}
\end{equation} 
and $\sigma$ are permutations of $\{1, ... , n\}$, and $\seq$ is the equilibrium entropy density. {The contribution we get from a permutation $\sigma$ at time $t$ on the right-hand side of \eqref{hol1} is equal to the contribution we get on the RHS of \eqref{hol2} from a configuration $\gamma(\sigma)$ where we first pair each $l_i$ with $r_{\sigma(i)}$, and then connect $l_i$ and $r_{\sigma(i)}$ if they are connectable. The set of $\gamma(\sigma)$ for all $\sigma$ at time $t$ is a subset of $\{\gamma\}(t)$ defined below \eqref{hol2}. We can obtain the full set $\{\gamma\}(t)$ if for each choice of $\sigma$, in addition to $\gamma(\sigma)$, we include configurations where any number of the connectable pairs of the type $\{ l_i, r_{\sigma(i)} \}$ are not connected. But disconnecting a pair of connected points in a configuration while leaving other points unchanged always increases the contribution from that configuration in \eqref{hol2}, so any element in $\{\gamma\}(t)$ which cannot be obtained as $\gamma(\sigma)$ for any $\sigma$ at time $t$  gives a larger contribution than some $\gamma(\sigma)$.  Thus, the minimum value in \eqref{hol2} is the same as the minimum in \eqref{hol1}.} 

{We can check that a minimal configuration in \eqref{hol2} only involves connections between adjacent endpoints. So we get the same result if we restrict the definition of connectable points below \eqref{hol2} to adjacent endpoints $l_i$ and $r_j$ such that $t>|l_i-r_j|/2$.} 

%\textcolor{red}{To see the equivalence between \eqref{hol1} and \eqref{hol2}, first note that at a given time $t$, a choice of permutation $\sigma$ in \eqref{hol1} can be mapped to an endpoint configuration $\gamma$ in the following way: $l_i$ is paired with $r_{\sigma(i)}$ if t> |l_i- r_{\sigma(i)}|/2$, and $l_i$ and $r_{\sigma(i)}$ are both unpaired points if $t< |l_i- r_{\sigma(i)}|/2$. Since every pair appearing in such a configuration is an allowed pair, such a $\gamma$ is one of the configurations appearing in $\{\gamma\}(t)$ defined below \eqref{hol2}, and the value we get from $\sigma$ on the RHS of \eqref{hol1} is the same as the value we get from $\gamma$ on the RHS of \eqref{hol2}.} 

 As the number of intervals increases, equation~\eqref{hol1} (or equivalently~\eqref{hol2}) gives rise to intricate patterns of time-dependence when the relative sizes of the intervals $|A_i|$ and their separations $|R_i|$ are varied.  
 It is remarkable that these patterns can be reproduced by the extremely simple underlying rules of \slcg\ and the \rgd.
We note, however, that while for the random unitary circuits, $S_2$ coincides with the von Neumann entropy 
 in the large $q$ limit, this is no indication that this is true for holographic systems in the large $c$ limit.\footnote{$S_2$ for a different configuration (two offset intervals in a thermal field double state) was calculated in~\cite{Asplund:2015eha} in a holographic system, and
 found to be different from the von Neumann entropy. $S_2$ in this setup can be calculated for random unitary circuits 
 and is found to agree with the von Neumann entropy, but not with $S_2$ of holographic systems.}
 So the entanglement spectrum of holographic systems cannot be fully approximated by random unitary circuits, 
 and while it is natural to expect that the \rgd\ should play some role in holographic systems, it cannot be the full story.

In~\cite{hong_mark}, it was shown using the strong subadditivity condition that the holographic expression~\eqref{hol1} in fact maximizes 
 the entanglement growth {for an arbitrary number of intervals} among all $(1+1)$-dimensional systems with a strict light cone. 
Thus we find that the \rgd~\eqref{rgd1} together with \slcg\ maximizes entanglement growth (recall that $S_2$ is upper-bounded by the von Neumann entropy).  In the two-interval case, this statement implies that any system with $S_2 (t)$ not equal to the holographic result must have $S_2(t)$ smaller than \eqref{unj0} and \eqref{unj}. Using  ~\eqref{myu0}, this implies that
 \be \label{iyr}
 G (A, B; t) \geq G_{\text{RVD}} (A, B; t) 
 \ee
 where  $G_{\text{RVD}} (A, B; t)$ is  the \rgd\ (RVD) expression~\eqref{unen}.

\iffalse 
 \ben 
 
 \item The \rgd~\eqref{rgd1}  also underlies holographic CFTs. 
 
\item \textcolor{red}{The evolution of holographic CFTs after a global quench can be approximated by a random unitary circuit.} 
 \item \textcolor{red}{For holographic systems after a global quench, in the large central charge and scaling limit, 
 all the Renyi and von Neumann entropies coincide as in the random circuit in the large $q$ limit.} 
 \een
\fi

\section{Void formation in two Clifford circuit models} \label{sec:circ}

As contrasts to the \rgd,  we now consider the void formation structure of two other examples of unitary circuits:
(i) a free propagating model in which entanglement 
can only be spread, but not created, which may thus be considered a proxy for free theories; 
(ii)  a circuit built from perfect tensors, which may be considered a model for interacting integrable systems, as while it can generate entanglement in certain initial product states, like all Clifford circuits it does not lead to growth of operator entanglement, and also does not have the form of the out-of-time-ordered correlator (OTOC) expected in chaotic systems \cite{frank, nahum1}.  Both models are special examples of Clifford circuits~\cite{c1,c2,c3}, a class of unitary circuits where under time evolution, a basis operator $\sO_\al$ transitions to another basis operator.

%\subsection{General description} 

More explicitly, consider a unitary circuit  defined by~\eqref{11}--\eqref{12} and Fig.~\ref{fig:circuit_structure},
where now each $U^{i,i+1}_t = \tilde{U}$, and $\tilde{U}$ is a (fixed) unitary matrix that evolves each basis operator 
in $\sG_i \otimes \sG_{i+1}$ to some  basis operator.  Under this time-evolution, for a given $\sO_{\alpha}$, $c_{\alpha}^{\; \beta}(t) = 1$ for a single $\beta$, and $c_{\alpha}^{\; \beta'}(t) = 0$ for all $\beta'\neq \beta$. Since the evolution is unitary, there is a one-to-one mapping from initial operators $\sO_{\alpha}$ to final operators $\sO_{\beta}$. 

To study entanglement growth, now instead of generic homogeneous pure product states, 
we have to consider a more restricted set, as  these Clifford circuit models do not generate entanglement in an arbitrary initial pure product state. Furthermore, initial basis operators in the entire system can in general both grow and decrease in size under the action of the circuit $U$. The set of initial states we look at are again of the form 
\begin{equation}
    \rho_0 = \frac{1}{q^L} \sum_{a \in I} \sO_{\alpha}
\label{general_initial}
\end{equation} 
where the set $I$ consists of 
$q^L$ basis operators, which are in general no longer 
 just tensor products of powers of $Z_i$. In each of the models, we will choose $\rho_0$ such that the end points of all basis operators in the associated $I$ 
move outwards with $v=1$ under time-evolution.
 
As before, the entanglement growth can be obtained from~\eqref{dens1}--\eqref{gfu}. Note that for Clifford circuits the off-diagonal terms in~\eqref{eej} vanish identically due to the one-to-one mapping between initial and final basis operators. The \slcg\ of operators in $I$ again implies that $S_2^{(A)}(t)$ is given by \eqref{heb} for a single interval.

In Clifford circuits,  we cannot have a void distribution like in~\eqref{rgd1} for individual initial operators $\sO_{\alpha}$, as the probability of going to any final operator is either $0$ or $1$. Thus the functions $N_A(t)$, $N(A, B;t)$ and $G(A, B;t)$ defined in~\eqref{ejn}--\eqref{gfu} are now the numbers~(rather than the expectation values of the numbers) of initial operators in $I$ satisfying a given property. It is instructive to contrast void formation functions $G(A, B;t)$
in these models with those from the random void distribution, and see how they lead to different entanglement growth. 
 
 Before discussing the models explicitly, here we summarize some common features, which are also shared by random unitary circuits in the large $q$ limit:
\begin{enumerate}
 
    \item While their void formation structure is very different from the \rgd, we will see the corresponding void formation functions nevertheless satisfy  the unitarity constraints~\eqref{heo} and~\eqref{ot1}--\eqref{ot3}.    
   
    \item The unitarity constraint~\eqref{heo}  
 is satisfied in the following way. At $t>|A|/2$, take any of the $q^{2t-|A|}$ basis operators in $I \cap X(A)$, say $\tilde{\sO}$, where $X(A)$ is the region of length $2t-|A|$ in the center of $J(A)$ (shown in Fig.~\ref{fig:regA}).  Defining $\tilde I$ as the set of all initial basis operators in $I \cap J(A)$ which are equal to $\tilde \sO$ in $X(A)$, one finds that 
\begin{equation}
    \sum_{\alpha \in \tilde I, ~ \beta \text{ trivial in } A} \coeff = 1  \ .
    \label{X_statement} 
\end{equation}
In the Clifford circuits, we can interpret this as the fact that any choice of initial operator within the region $X(A)$ is consistent with evolving to a final operator which is equal to the identity in $A$. Since the total number of operators in $I\cap X(A)$ is equal to $q^{2t-|A|} = G(A, J(A); t)$, this also means that when we fix the part  of the initial operator within $X(A)$, the initial operator in the entire region $J(A)$ which can evolve to the identity is fully determined. 

It is tempting to speculate that given \slcg, equation~\eqref{X_statement} is  true in all unitary systems.  

\end{enumerate}

%Below we discuss explicitly two Clifford circuit models respectively in Sec.~\ref{sec:free_prop} and \ref{sec:P_tensor}.  

%and compare results from those models with those from \rgd\ in Sec.~\ref{sec:comparison}.

%We introduce the and show how they satisfy the above propertiesbelow. 

% As we show in section  \ref{sec:comparison}, $S_2(t)$ for two intervals is determined by $N(\bar{A}, B, t)$ for a single interval $A$. So we compare $N(\bar{A}, B, t)$ for a single interval $A$ in the Clifford circuit models to \eqref{unen} for the random void distribution, and look at the resulting differences in $S_2(t)$ for two intervals among the three models. 

%\textcolor{red}{Include general comment about chaotic/integrable nature of Clifford circuits based on behaviour of operator entanglement and OTOCs?}

%.

\subsection{Free propagation model} 
\label{sec:free_prop}
In this model, the two-site unitary matrices in the circuit are given by~\cite{hong_mark}
\be 
\tilde{U}_{a_1 a_2, b_1 b_2} = \de_{a_1 b_2} \de_{a_2 b_1}
\ee
which is a discrete version of the quasiparticle models for entanglement growth proposed in~\cite{calabrese}.
$\tilde U$ takes a product state to a product state at all times, but can spread the entanglement to large distances
if we consider an initial state
 \begin{equation} 
\begin{gathered}
\ket{\psi_0} = ...\otimes (\frac{1}{\sqrt{q}}\sum_{n=0}^{q-1}\ket{n}_0\otimes \ket{n}_1) \otimes  (\frac{1}{\sqrt{q}}\sum_{n'=0}^{q-1}\ket{n'}_2\otimes \ket{n'}_3) ... 
\end{gathered}
\label{ent_state}
\end{equation}
which has short-range entanglement between adjacent pairs of sites.

The evolution of basis operators\footnote{We use the following conventions to avoid complications due to lattice effects which are not relevant in the continuum limit. All spatial regions we consider have their left endpoint at an even site and their right endpoint at an odd site. We always consider times at which an odd number of layers of unitaries have been applied.} 
in this model has a simple form: the basis operators at different sites evolve independently from each other, and all operators that are initially at an even (odd) site move to the right (left) at speed 1, so that at an odd time $t$, for an initial operator 
$\sO \equiv \otimes_i O_i$, 
\begin{equation} 
\sO(t) = \otimes_{i=\text{even}} O_{i+t} \otimes_{i=\text{odd}} O_{i-t}
\label{op_ev_quasi}
\end{equation}

\begin{figure}[h]
    \centering
    \includegraphics[width=10cm]{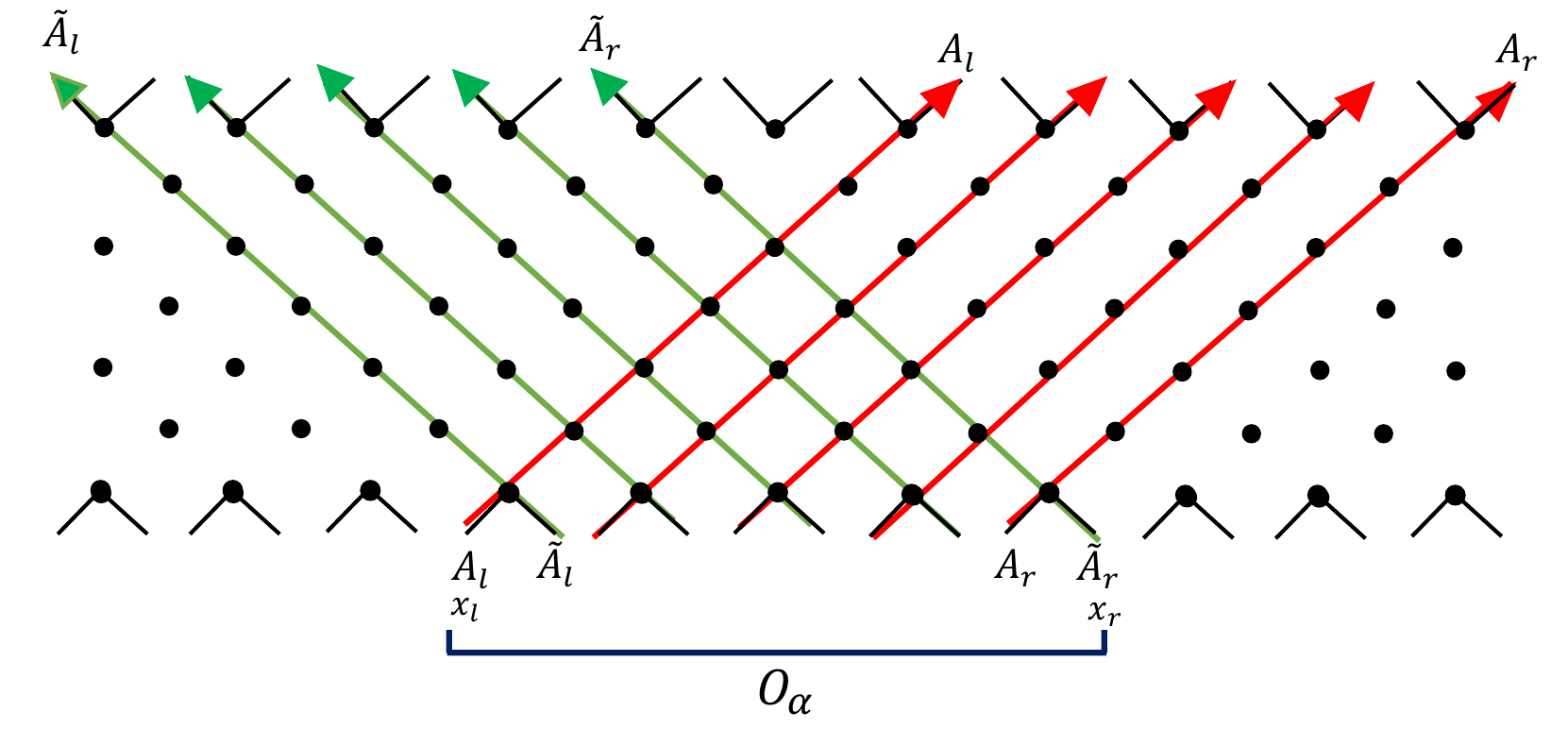}
    \caption{Growth of an operator $\sO_{\alpha}$ in $I$. Any operator in $I$  has its left endpoint at an even site $x_l$ and its right endpoint at an odd site $x_r$. The evolution of operators follows~\eqref{op_ev_quasi} with operators at even (odd) sites moving right (left), which are shown respectively using red and green.
 At time $t$, the left and right endpoints are at $x_l-(t-1)$ and  $x_r+(t-1)$ respectively (for sufficiently large times we have $t-1\approx t$). The resulting operator has a central void of length $2t-L_i$, where $L_i$ is the length of the initial operator, as well as other voids of length $1$. {The number of sites with non-trivial operators remains fixed as a function of time.}}
    \label{fig:growth_qp}
\end{figure}

 The density matrix for~\eqref{ent_state} is of the form \eqref{general_initial}, with $I$ the set of operators which can have any basis operator $O_i$ at an even site $i$, but at site $i+1$ have some fixed $\tilde{O}_{i+1}$ determined by $O_i$. For example, if $O_i = X^m Z^n$, then $\tilde{O}_{i+1} = X^m Z^{q-n}$. 
The time evolution of basis operators in $I$ can be readily obtained from~\eqref{op_ev_quasi},  see Fig.~\ref{fig:growth_qp} for an illustration. 

\begin{figure}[!h]
    \centering
    \includegraphics[width=10cm]{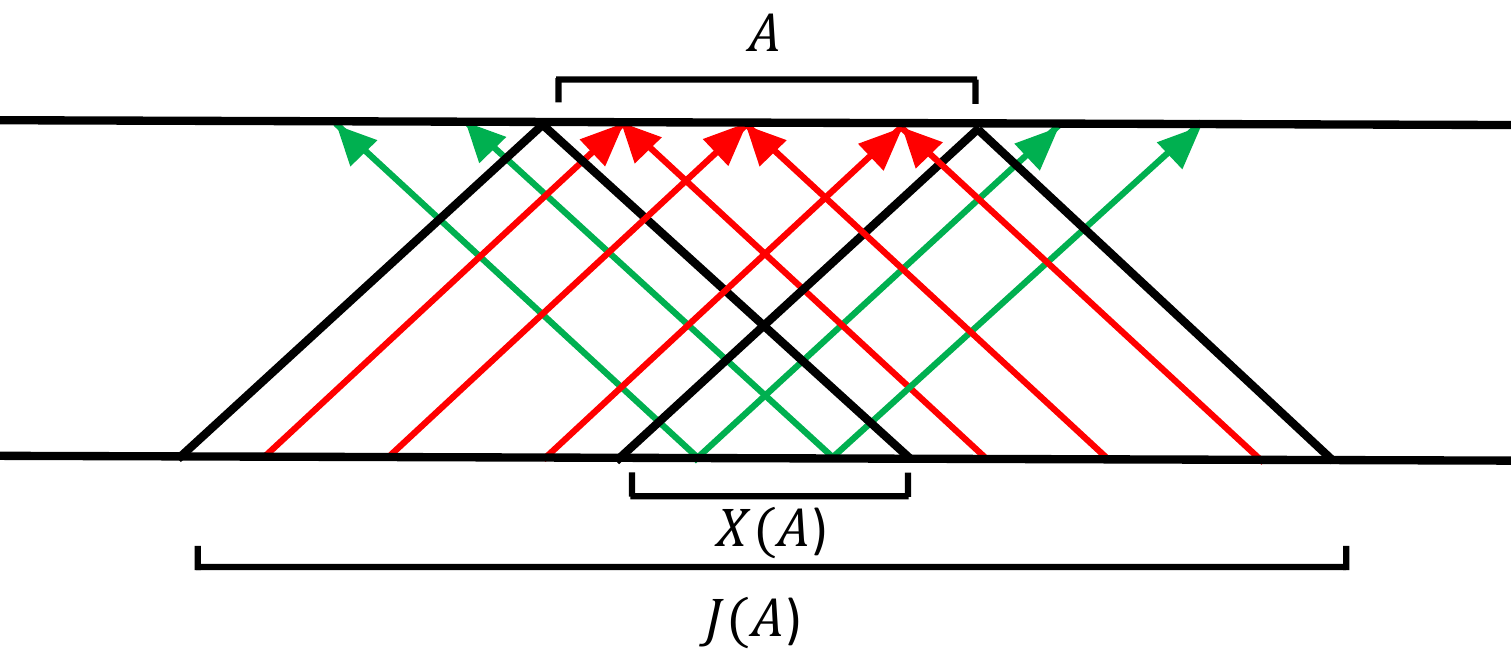}
    \caption{All
    operators from the region $X(A)$ propagate to $\bar{A}$. {Any operator non-trivial  in\\$J(A)-X(A)$ has a part that propagates to $A$.}}
    \label{fig:AXJ}
\end{figure}

In Fig.~\ref{fig:AXJ}, we see that all initial operators from sites in $X(A)$ propagate to $\bar{A}$ at time $t$, while operators on sites in the region $J(A)-X(A)$  propagate to $A$. An operator $\sO_{\alpha}$ in $I$ becomes trivial in $A$ if and only if it is of the form $\sO_{\alpha}= \tilde{O}_{X(A)}\otimes \mathbf{1}_{J(A)-X(A)}$, where $\tilde{O}_{X(A)}$ can be any operator in $I \cap X(A)$. Thus the statements \eqref{heo} and \eqref{X_statement} are both true in this model. 

Now look at the form of $G(A, B; t)$. Since any initial operator in $J(A)\cap I$ that becomes trivial in $A$ must be contained within $X(A)$, the number of basis operators in $I$ contained in a subset $B$ of $J(A)$ that develop a void in $A$ is given by 
\begin{equation}
    G_{\rm free} (A, B; t) = q^{|B\cap X(A)|} \ .
    \label{int_gap}
\end{equation}
Note that this is very different from the form~\eqref{unen}  from the random void distribution. In particular, while \eqref{unen} depends only on the length of $B$ and not on its position in $J(A)$, \eqref{int_gap} is sensitively dependent on the position of $B$ in $J(A)$. Moreover, equation~\eqref{int_gap} can be greater than $1$ when $|B|< 2|A|$, if $B$ has some overlap with $X(A)$. When applied to the entanglement entropy for two intervals using~\eqref{myu0}, such behavior can lead to 
$S_2^{(A)} (t)$ behaving non-monotonically in time\footnote{which was well known in the context of the quasiparticle model~\cite{Balasubramanian:2011at,Asplund:2013zba,Leichenauer:2015xra}}, whereas~\eqref{unj0}--\eqref{unj} are non-decreasing. 

%Fig.~\ref{fig:ent_2} gives an example. 
\iffalse
\begin{figure}
    \centering
    \includegraphics[width=8cm]{ent_2.pdf}
    \caption{An example for which $G(A, B; t)$ is nonzero for $|B|< 2|A|$. When applying~\eqref{myu0}, this can lead to $S_2^{(A)} (t)_{\text{free}}$ behaving non-monotonically in time. \HL{do we need $A_1, A_2$? Move earlier?}} 
    \label{fig:ent_2}
\end{figure}
\fi

One can also see that, as anticipated from~\eqref{iyr}, the expression~\eqref{int_gap} is always greater than~\eqref{unen}
\be \label{kju}
G_{\text{free}} (A, B; t) \geq G_{\text{RVD}} (A, B; t), \quad \forall A, B  \ .
\ee
Equation~\eqref{kju} is  equivalent to $|B\cap X(A)|\geq |B|-2|A|$, which is always true as $|B| - |B\cap X(A)| \leq  J(A) - X(A) = 2 |A|$ (see Fig.~\ref{fig:AXJ}).  {It can also be checked the constraints \eqref{ot1}--\eqref{ot3} obtained from unitarity are satisfied in this model.}

\iffalse
\begin{figure}
\centering 
\includegraphics[width=14cm]{quasiparticle_B.pdf}
\caption{If $B$ is as shown in figure (a), then $|B|\leq |B\cap X(A)| + |A|$, as the part of $B$ outside $X(A)$ is contained within a region of total length $|A|$. The same is true if the part of $B$ outside $X(A)$ has overlap with the region to the right of $X(A)$, but not with the region to the left. If $B$ is as shown in figure (b), then $|B|\leq |B\cap X(A)| + 2|A|$, as the two parts of $B$ outside $X(A)$ are contained regions which each have length $|A|$.  }
\label{fig:B_position}
\end{figure} 
\fi

%Note that in this simple model the calculation of the growth of $S_2$ using the propagation of operators exactly parallels the calculation of the growth of $S_{\text{VN}}$ in \cite{hong_mark}, and $S_2= S_{\text{VN}}$. \textcolor{red}{Would it be useful to show this more explicitly?}    

\subsection{Perfect tensor model}\label{sec:P_tensor}

In this model, which was previously considered in~\cite{qi}, the Hilbert space at each site has dimension $3$ ($q$ below should be understood as being $3$), with a
basis $\{\ket{0}, \ket{1}, \ket{2}\}$, and {$\tilde{U}^{\dagger}$} acts on
the Hilbert space as:
\be\label{jhy}
\tilde{U}^{\dagger} \ket{i} \otimes \ket{j} = \ket{-i - j} \otimes \ket{j-i}  ,
\ee
with addition defined modulo 3.
 $\tilde{U}$ is a perfect tensor, that is,  any balanced bipartition of its indices
into inputs and outputs gives a unitary transformation.  The perfect tensor model does not generate entanglement in every initial pure product state: for example,  states $...\ket{0}\otimes \ket{0}...$ and $...\ket{\psi_3}\otimes \ket{\psi_3}...$  remain invariant under the action of $U$. 
We will consider an initial state of the form  
\begin{equation}
    \ket{\psi} = ...\otimes \ket{\psi_3}_0\otimes \ket{0}_1 \otimes  \ket{\psi_3}_2\otimes \ket{0}_3 ...
\label{initial_perfect}
\end{equation}
where $\ket{\psi_3}\equiv \frac{1}{\sqrt{3}}\sum_{k=0}^{2}\ket{k}$, for which $I$ is the set of basis operators with powers of $X$ on even sites and powers of $Z$ on odd sites. 

From~\eqref{jhy} one finds that acting on operators in the basis~\eqref{obd},  $\tilde U$ sends any basis operator on two sites to another basis operator,
  \begin{equation} 
    \tilde{U}^{\dagger}  (b_1 \otimes b_2) \tilde{U}= b'_1 \otimes b'_2 ,
    \label{B1B2}
    \end{equation}
and has the following properties: 

\ben 

\item  It takes any operator with a power of $X$ on site $i$ and a power of $Z$ on site $i+1$ to an operator which is non-trivial on both $i$ and $i+1$. 

\item It takes  any basis operator non-trivial on a single site to a basis operator non-trivial on both sites.

\item  If we know any two of $b_1, b_2, b'_1, b'_2$, the other two are fully determined. 

\item For $b_1 \otimes b_2$ of the form $X^m \otimes Z^n$, the set of $b_1'$ runs over all one-site basis operators and 
if we fix $b_1'$, then $b_2'$, $b_1$ and $b_2$ are uniquely determined (and similarly if we fix $b_2'$, $b_1'$, $b_1$ and $b_2$ are uniquely determined). 

\een

\begin{figure}[!h]
    \centering
    \includegraphics[width=8cm]{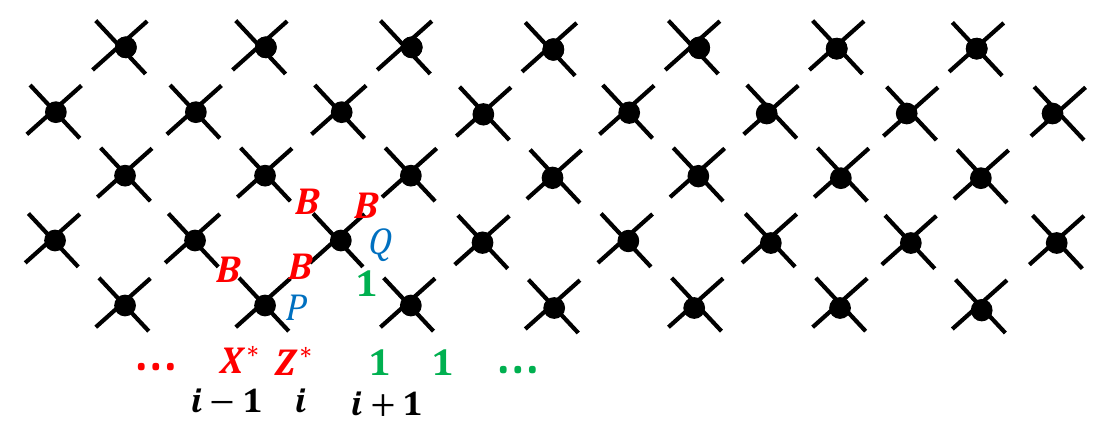}
    \caption{Sharp light-cone growth of operators in the perfect tensor model.  The unitary $\tilde{U}$ at point $P$ in the circuit acts on operators at $i-1$, $i$, and gives some operator which is nontrivial on both $i-1$ and $i$. In the figure, $X^{\ast}, Z^{\ast}$ refer to any powers of $X$ and $Z$, and $B$ to any non-trivial operator. 
 $\tilde{U}$ at $Q$ acts on an operator which is non-trivial on site $i$ and trivial on site $i+1$, giving an operator which is non-trivial on both $i$ and $i+1$. By  repeatedly using the fact that operators nontrivial on a single site evolve to operators non-trivial on two sites, we see that the right endpoint moves to $i+(t-1)$ at time $t$.}
    \label{fig:operator_pt}
\end{figure}

From items (1) and (2), we see that all operators in $I$ grow outwards with speed 1, as illustrated in Fig.~\ref{fig:operator_pt}. Thus we have the same form of $S^{(A)}_2(t)$ for a single interval $A$ as we found in the random circuit and free propagation models with the initial states we considered there. Note that~\eqref{initial_perfect} is a product state, so unlike the free propagation model, the perfect tensor model can generate entanglement in an initial pure product state. 

\begin{figure}[!h]
    \centering
    \includegraphics[width=10cm]{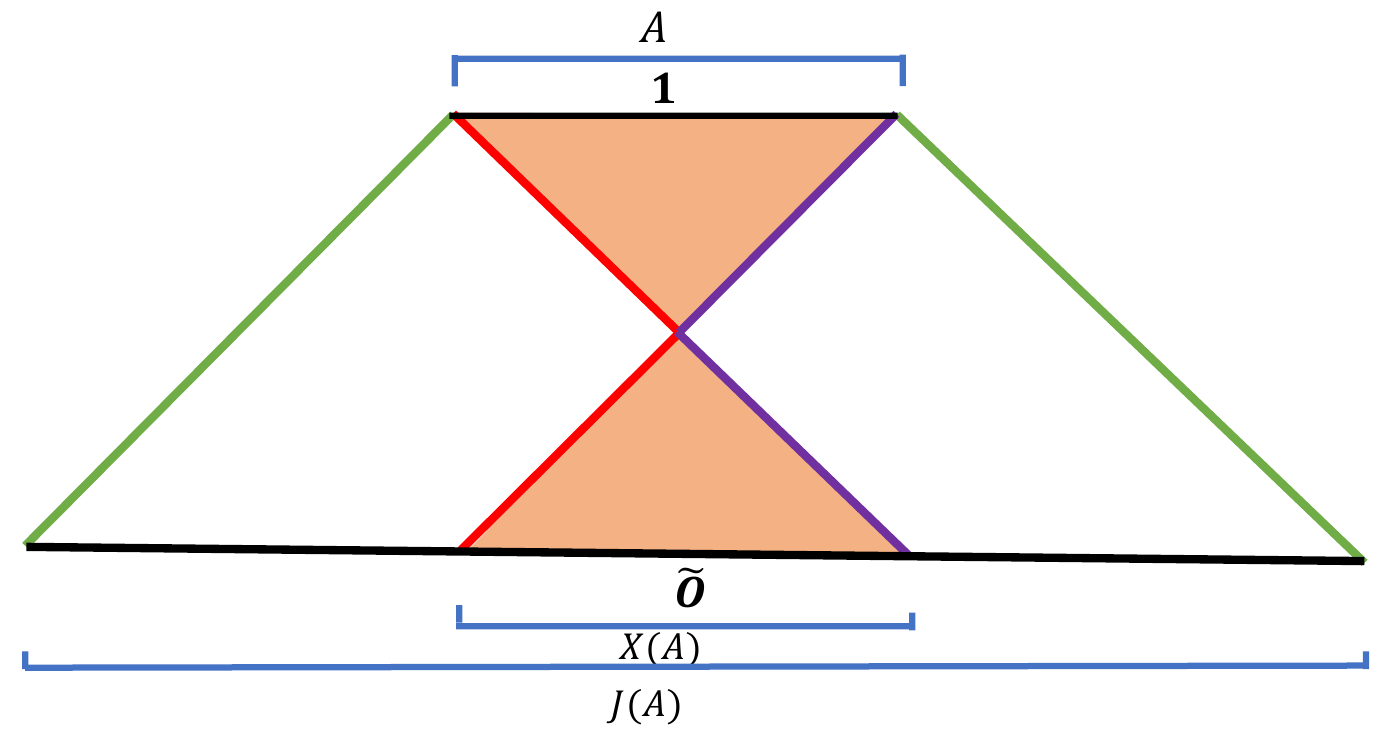}
    \caption{From causality the {operators} in the past domain dependence of $A$ 
    and future domain dependence of $X(A)$ are fixed by operators in $A$ and $X(A)$ respectively. From the two ``Cauchy surfaces'', shown in red and purple, 
    which bound the regions of determined operators, by repeatedly using items (3) and (4)  one can uniquely determine the 
    all other operators in the portion of the circuits between the green lines, including the remaining parts of the initial operator in $J(A)$.}
    \label{fig:perfect_AXJ}
\end{figure}

From items (3) and (4), one can show that for any basis operator $\sO_\al = P_{X(A)} \otimes Q_{J(A)-X(A)} \in J(A) \cap I$, 
for  $\sO_\alpha(t)$ to be $\mathbf{1}$ in the region $A$, $P_{X(A)}$ can be any operator in $I\cap X(A)$, and if we fix $P_{X (A)}$, then ${Q}_{J(A)-X}$ is uniquely fixed.  The basic idea is illustrated in Fig.~\ref{fig:perfect_AXJ}. 
This implies~\eqref{X_statement}, and also implies \eqref{heo} as $G(A, J(A), t)$ is equal to the number of basis operators in $I\cap X(A)$, which is $q^{2t-|A|}$. 
Note that an initial operator that becomes trivial in $A$ at time $t$ will in general be non-trivial in the region $J(A)-X(A)$, unlike in the free propagation model. 
    
\begin{figure}[!h]
\begin{center}
\includegraphics[width=16cm]{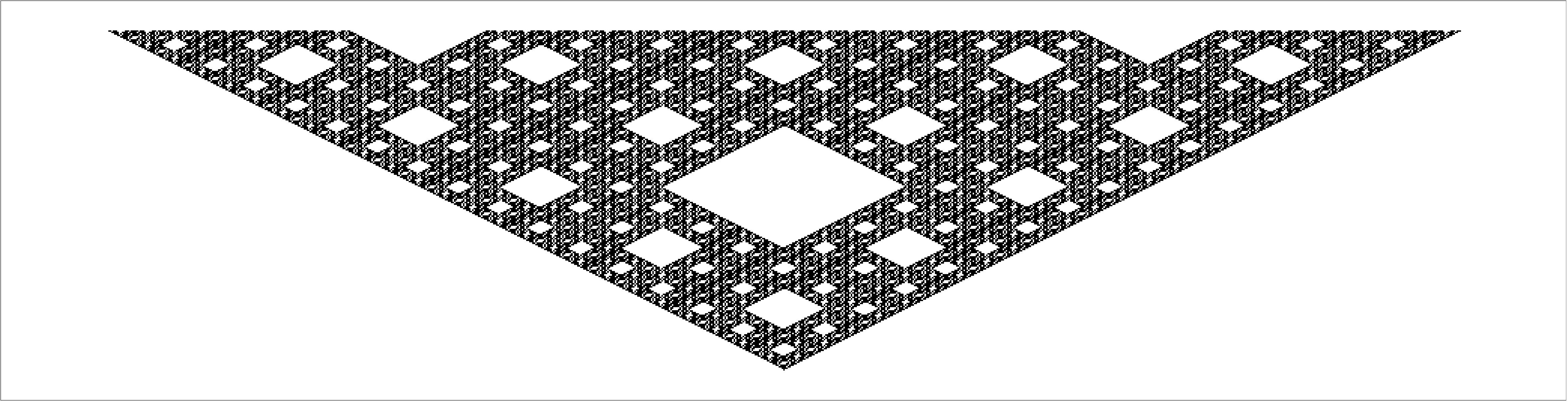}
\includegraphics[width=16cm]{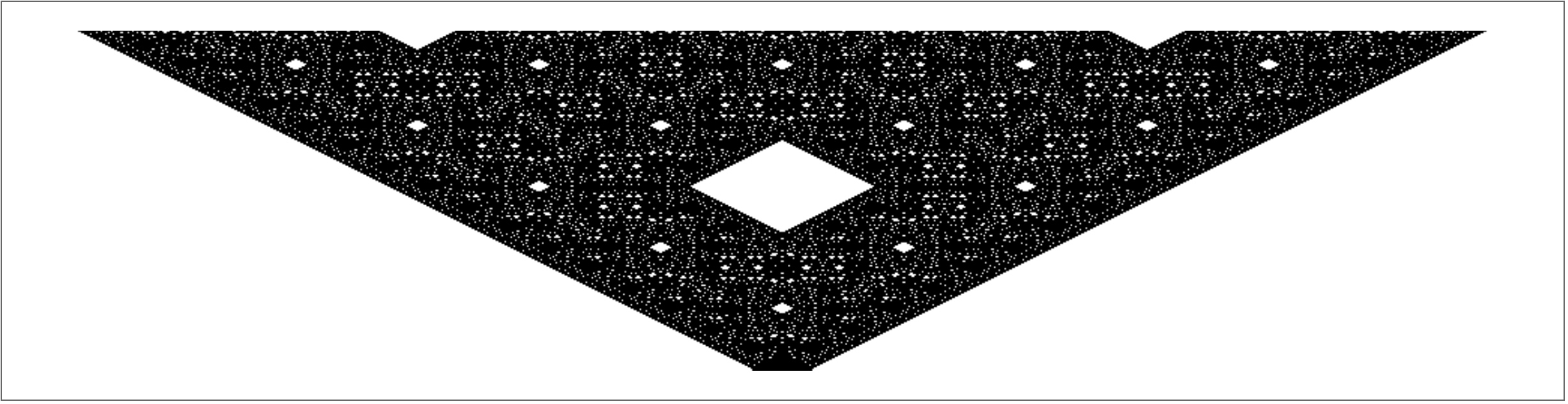}  
\caption{In the top figure, we show the time-evolution of an initial operator from $I$ which is non-trivial on one site in the perfect tensor model. In the bottom figure, we show the time-evolution of an operator non-trivial on 40 sites. Time increases in the upward direction, and the time-evolved operator has non-trivial support on black sites, and the identity operator on white sites. It is evident from the examples that for fixed time, fewer voids are formed in the case where the initial operator is bigger.}
\label{fig:perfect_ev}
\end{center}
\end{figure}

Further differences from the free propagation model can be seen by looking at the time-evolution of single initial operators in Fig.~\ref{fig:perfect_ev}. The void formation in this model has a fractal structure, similar to the fractal Clifford circuits discussed in \cite{c2,c3}.\footnote{In \cite{c3}, all Clifford circuits for $q=2$ are classified into three types: periodic, glider and fractal. The latter two types cannot leave any pure translation-invariant stabilizer state invariant. In the perfect tensor model here with $q=3$, the operator evolution has a structure similar to fractal Clifford circuits, but as mentioned earlier this model leaves some pure  translation-invariant stabilizer states such as $\ket{0}\otimes ... \otimes \ket{0}$ invariant.} The number of non-trivial operators in $J^+(O_{\alpha})$ grows unboundedly with time, in contrast to the situation in the free propagation model shown in Fig.~\ref{fig:growth_qp}.
 
\begin{figure}[!h] 
\centering 
\includegraphics[width=6cm]{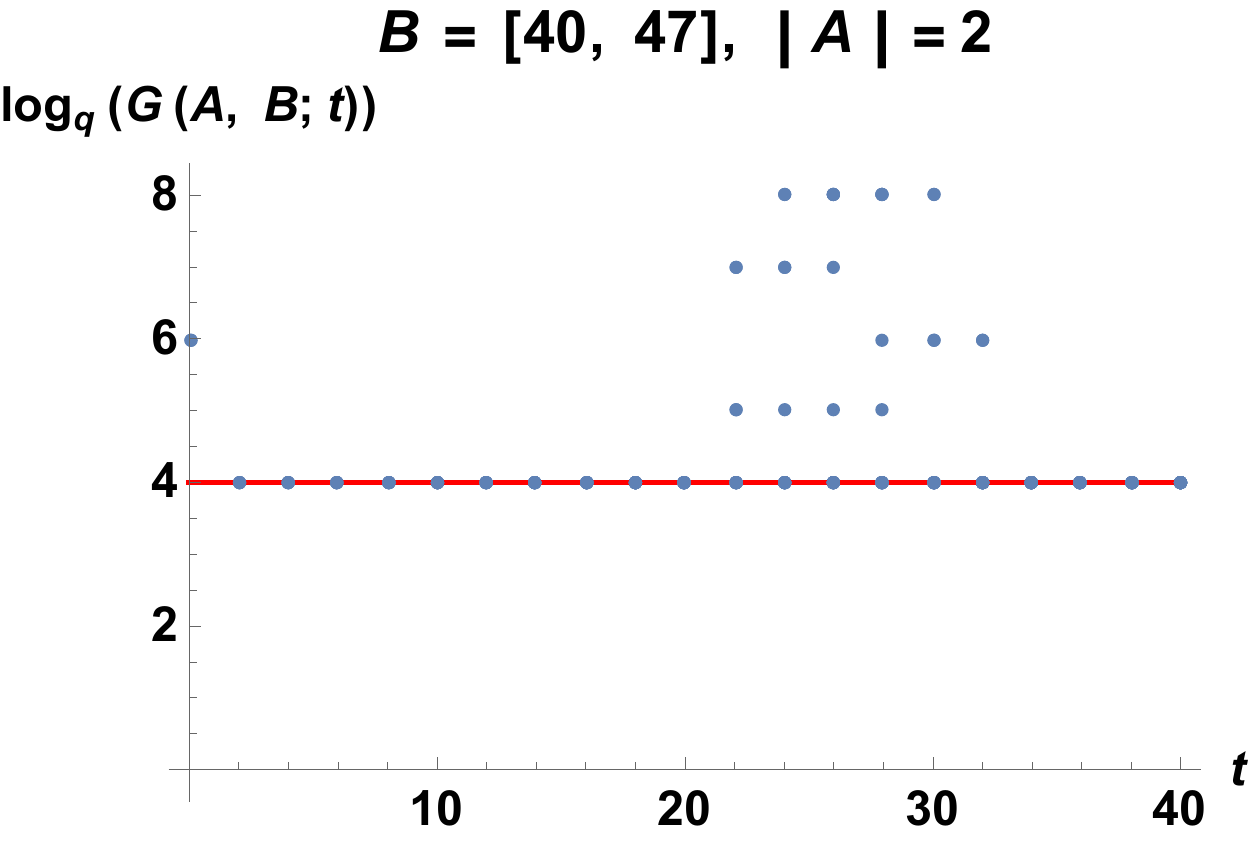} ~~~
\includegraphics[width=6cm]{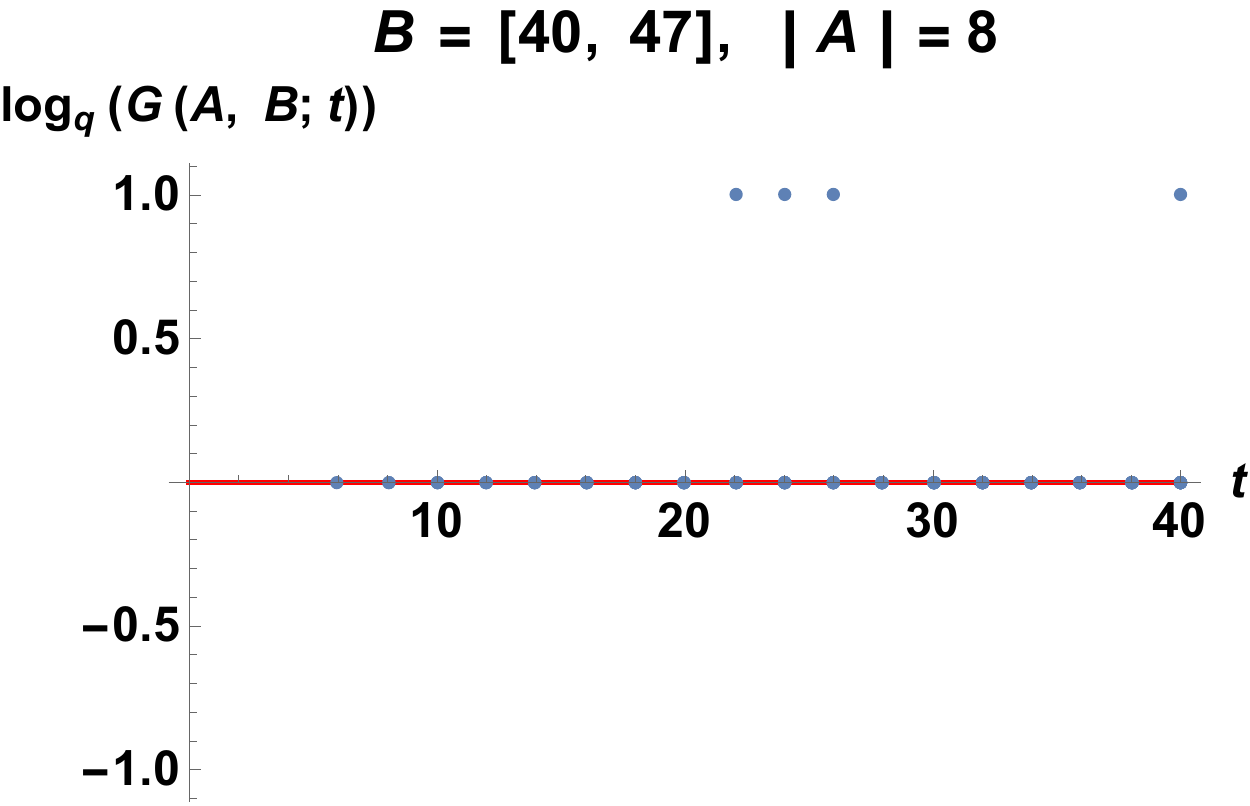}
\caption{We consider a system with $L=80$, and sites labelled from $x=0$ to $x=79$. We fix the region $B = [40, 47]$, and look at $G(A, B;t)$ for different choices of $A$. We consider intervals with $|A|=2<|B|/2$ on the left, and $|A|=8> |B|/2$ on the right, and in both cases we sample over a large number of intervals $A$ and show $\log_q(G(A, B;t))$ as a function of time in a single plot with the blue data points. The value expected from the random void distribution~\eqref{unen} is $4$ in the first case, and $0$ in the second case. We see clearly that  $G(A, B;t) \geq G_{\text{RVD}}(A, B;t)$, and there are  cases where $G(A, B;t) >G_{\text{RVD}}(A, B;t)$.}
\label{fig:perfect_GABT}
\end{figure}

We do not have a closed form for the void formation function $G(A, B; t)$ for a general region $B$, but one can readily check 
in examples that~\eqref{iyr} is satisfied. See Fig.~\ref{fig:perfect_GABT}. We do not have a general  derivation of the constraints \eqref{ot1}-\eqref{ot3} in this model, but we have checked in a number of examples that they are obeyed. 

\section{Conclusions and discussion} \label{sec:conc}

In this paper, we examined the implications of void formation in operator evolution 
for entanglement growth, and showed that it plays an important role 
in maintaining unitarity of entanglement growth and generation of mutual information and multi-partite entanglement. 
We showed that the void formation probability for generic operators in random unitarity circuits is given by the 
 \rgd~\eqref{rgd1}. We also showed that the intricate time-dependence of holographic entanglement entropies for an arbitrary number of intervals after a global quench can be understood as a consequence of the very simple rules of \slcg\ and the \rgd.

%\HL{This agreement with the holographic results can be viewed as support for the conjecture that generic operators in a quantum chaotic system obey the \rgd.}  

%\textcolor{red}{The fact that we can found sets of operators with sharp light-cone growth in the chaotic as well as integrable models we considered, 

Furthermore, we found sharp differences between the void formation properties of random unitary circuits and the non-chaotic circuit models we studied, which suggests that void formation can be used to characterize differences in the operator evolution of chaotic and integrable systems which are not captured by the movement of the operator endpoints alone~\cite{Kem}.
It is also an interesting question whether void distribution can be used to distinguish between different classes of chaotic systems. 
For example, it is conceivable that the \rgd\ may only apply to highly chaotic systems.

In our discussion, for simplicity of presentation we have restricted to systems with \slcg\ in the evolution of operators.
In general systems, the fronts of operator growth should follow a distribution. For example, in random unitary circuits 
at finite $q$, the evolution of an operator exhibits a diffusive front around a butterfly velocity $v_B$ which is smaller than the 
lightcone velocity $v_c$~\cite{nahum1,frank}. Our discussion, including the \rgd, can be generalized to these situations, although the story is technically more complicated and will be presented elsewhere.  

In our discussion we have focused on the second Renyi entropy, which can be conveniently expressed in terms of 
the expected number of operators that develop a void in a certain region. It would be interesting to explore 
the implications of void formation in operator growth for higher Renyi and von Neumann entropies\footnote{although in some cases, like random unitary circuits in the large $q$ limit, all the entropies are the same.}, as well as 
whether the unitarity of these quantities imposes further constraints on void formation.  It is possible these quantities will involve 
other aspects of void formation, and not just the squared absolute values of the operator-spreading coefficients. 

While we have restricted to one spatial dimension, our discussion can be immediately generalized to higher dimensions. 
In one dimension, a process of void formation separates both the original operator and the full space into disconnected parts. 
Thus it simultaneously creates ``holes'' in an operator and breaks it into disjoint pieces. 
This is not true in higher dimensions, where  ``hole formation'' in an operator and 
breaking an operator apart are distinct void formation processes, as discussed in Fig.~\ref{fig:mul}. 
In particular, it is the latter type of process which contributes
to mutual information and multi-partite entanglement among disjoint regions. It is also a natural question whether the ``hole formation'' and ``breaking apart'' could follow different probability distributions in higher dimensions.

In this paper, we defined a void as a region of identity operators among regions of nontrivial support of an operator. 
This definition is only appropriate for a lattice system with a finite one-site Hilbert space at infinite temperature. 
For finite temperature or continuum systems, a mathematically rigorous definition is tricky. Operationally,  one can define a void as the part of an operator which is given by the equilibrium density operator.

It would be interesting to explore the implications of void formation for other observables to see how it affects their behavior in integrable and chaotic systems, and also to see if the non-unitarity in the absence of void formation manifests itself in other observables.
For example, one can show that in some simple models without void formation, the out-of-time-ordered correlation functions~(OTOCs)
will also violate unitarity, although the violation appears less dramatic than that in the entanglement entropy.  
We will leave the exploration of this question for elsewhere. 

It is important to see whether one can find other measures to characterize void formation besides than the probability functions we discussed in this paper. For example, how does one characterize the fractal structure of Fig.~\ref{fig:perfect_ev}? 
A related question is whether such fractal structure is generic among interacting integrable systems. 

%The discussion of
%could be helpful for this purpose. 

\vspace{0.2in}   \centerline{\bf{Acknowledgements}} \vspace{0.2in}
We would like to thank Netta Engelhardt, Paolo Glorioso, Aram Harrow, Lampros Lamprou, Sam Leutheusser, Raghu Mahajan, Juan Maldacena,  M\'ark Mezei, Adam Nahum,  Tibor Rakovszky, Jan Zaanen and Ying Zhao
 for discussions. 
This work is supported by the Office of High Energy Physics of U.S. Department of Energy under grant Contract Number  DE-SC0012567.

\appendix

%\section{Second Renyi entropy for multiple intervals in random unitary circuits}

\section{Derivations in the random unitary circuits} 
\label{app:part}

In this Appendix, we first briefly review the partition function method introduced in~\cite{nahum1,frank,nahum2}
for calculating various observables in random unitary circuits, and then use it to derive the \rgd~\eqref{rgd1} and the
time evolution of entanglement entropy. 

\subsection{Mapping to a classical Ising partition function} 

Consider the operator spreading coefficient $\coeff$ introduced in~\eqref{evop}, which can be written as a matrix element on four copies of the Hilbert space: 
\begin{equation} 
\begin{aligned}
\coeff &= \le| \frac{1}{q^L} \text{Tr}\le(\sO_{\beta}^{\dagger} U^{\dagger} \sO_{\alpha} U\ri) \ri|^2 \cr
& = {1 \ov q^{2L}} (\sO_\al)_{b_1 a_1} (\sO_\al^\da)_{d_1 c_1}  U_{a_1 a_2} U^*_{b_1 b_2} U_{c_1 c_2} U^*_{d_1 d_2} 
(\sO_\b^\da)_{a_1 b_1} (\sO_\b)_{c_1 d_2} \cr
& = \bra{{\sO_{\alpha}}_{\uparrow}}~ (U \otimes U^{\ast}\otimes U \otimes U^{\ast})~ \ket{{\sO_{\beta}}_{\uparrow}} \ .
\end{aligned}
\label{op_exp}
\end{equation} 
In the last line we have introduced, for any operator $\sO$ in the system,  ``up" and ``down" spin states on four copies of $\mathcal{H}$, 
\begin{equation} 
\braket{ abcd \mid \sO_{\uparrow}} = \frac{\sO^{\dagger}_{ab} \sO_{cd}}{\text{Tr}[\sO \sO^{\dagger}]},  ~~~~~ \braket{ abcd \mid \sO_{\downarrow}} = \frac{\sO^{\dagger}_{ad} \sO_{cb}}{\text{Tr}[\sO \sO^{\dagger}]}.
\label{op_state}
\end{equation}
In the case where $\sO= \mathbf{1}$, we use the notation $\ket{\uparrow} = \ket{\mathbf{1}_{\uparrow}}, \ket{\downarrow} = \ket{\mathbf{1}_{\downarrow}}$.

The time evolution operator $U$ for the entire system is a tensor product of random unitaries from the Haar ensemble applied at each time on pairs of sites, as shown in~\eqref{11}--\eqref{12} and Fig.~\ref{fig:circuit_structure}. As explained in \cite{nahum1,frank,nahum2}, after averaging over local unitaries with the Haar measure, one can express~\eqref{op_exp} 
as a partition function of classical Ising spins on a triangular lattice, as shown in Fig.~\ref{fig:coeff_lattice},  with the following specifications: 

\begin{figure}[!h]
    \centering
   \includegraphics[width=12cm]{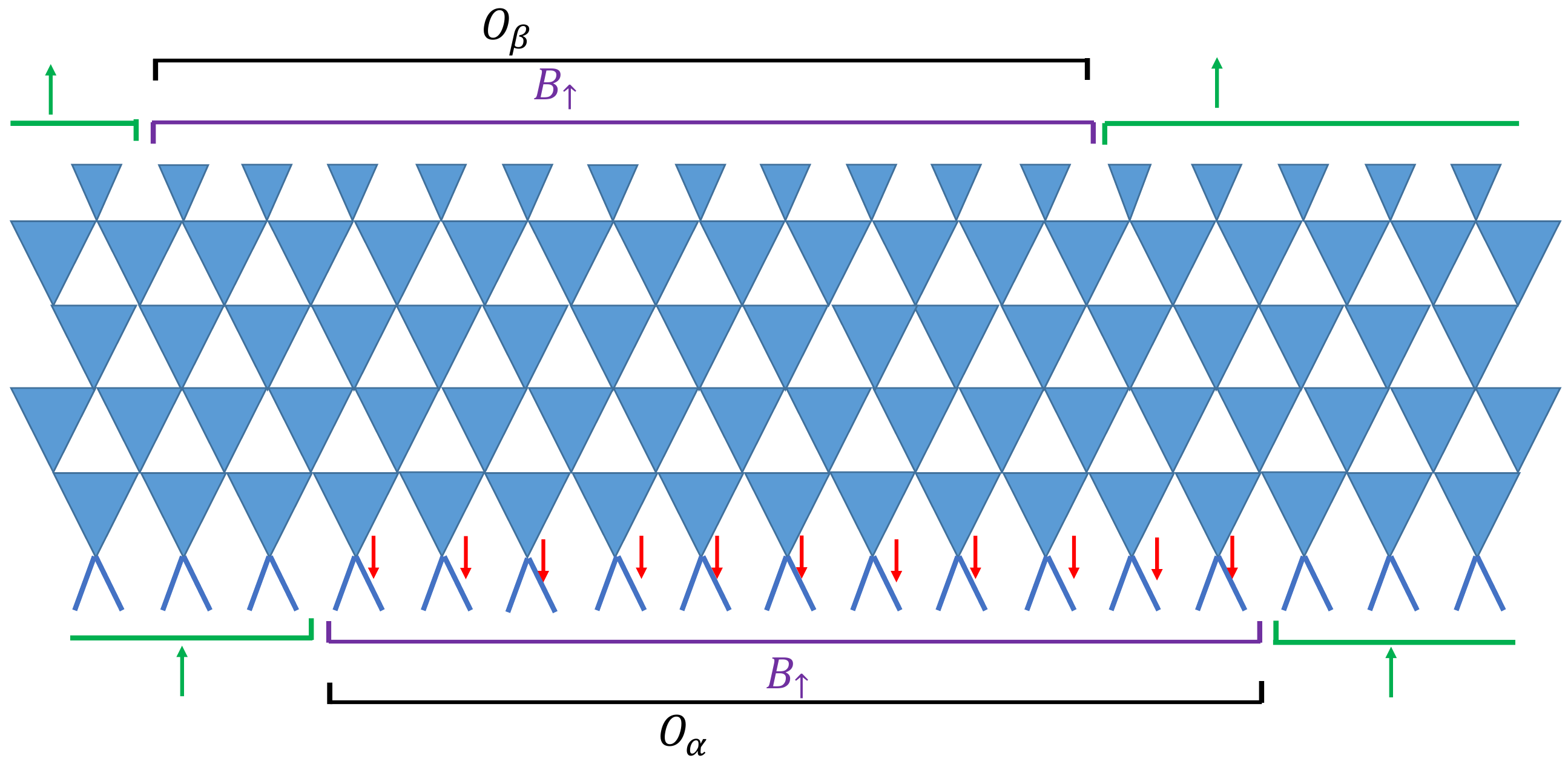}
    \caption{We show the lattice for the partition function corresponding to \eqref{op_exp} for a particular choice of $O_{\alpha}$ and $O_{\beta}$, both without voids, and $t=5$. The boundary conditions, and the spins in the bulk fixed by the boundary conditions through the rules in Fig.~\ref{fig:bottom_rules}, are shown explicitly. The remaining spins can be either up or down.}
    \label{fig:coeff_lattice}
\end{figure}

\ben 

\item The top layer of the lattice corresponds to time $t$ and the bottom layer to $t=0$. They are determined respectively by 
$\ket{{\sO_{\beta}}_{\uparrow}}$ and $\ket{{\sO_{\al}}_{\uparrow}}$.  If $\sO_{\alpha} = \otimes_i O_i$, then the spins on the lower boundary are given by $O_{i \uparrow}$ at site $i$. {$O_{\beta}$ similarly fixes the spins on the top boundary.}  

\begin{figure}[!h]
    \centering
    \includegraphics[width=11cm]{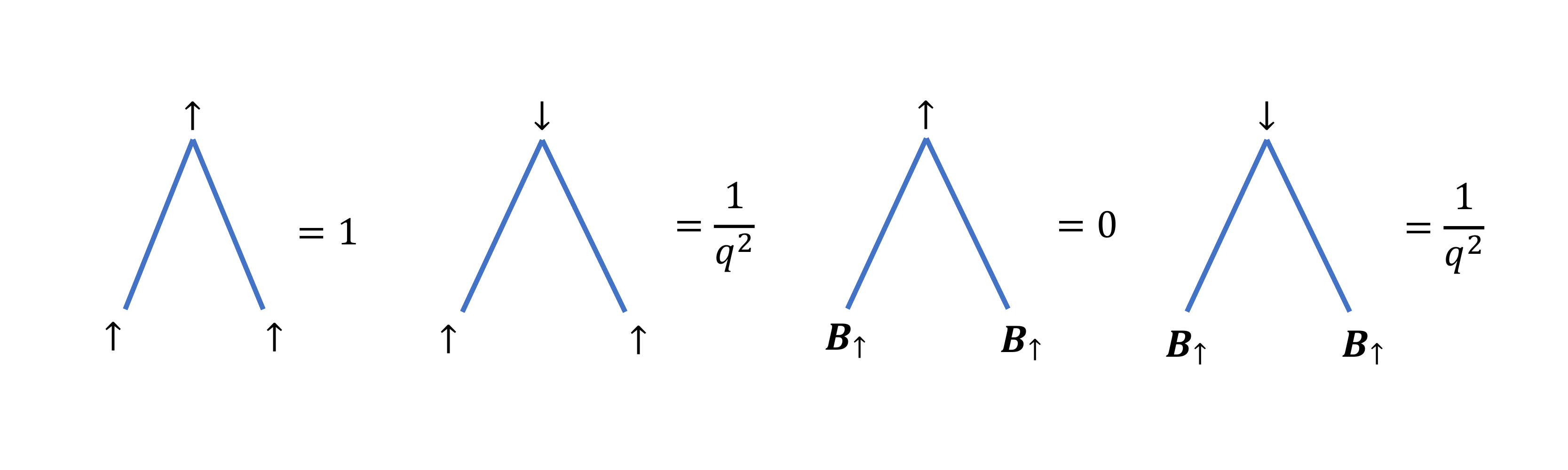}
    \caption{Rules for interactions along the bottom boundary. }
    \label{fig:bottom_rules}
\end{figure}

\begin{figure}[!h]
    \centering
    \includegraphics[width=11cm]{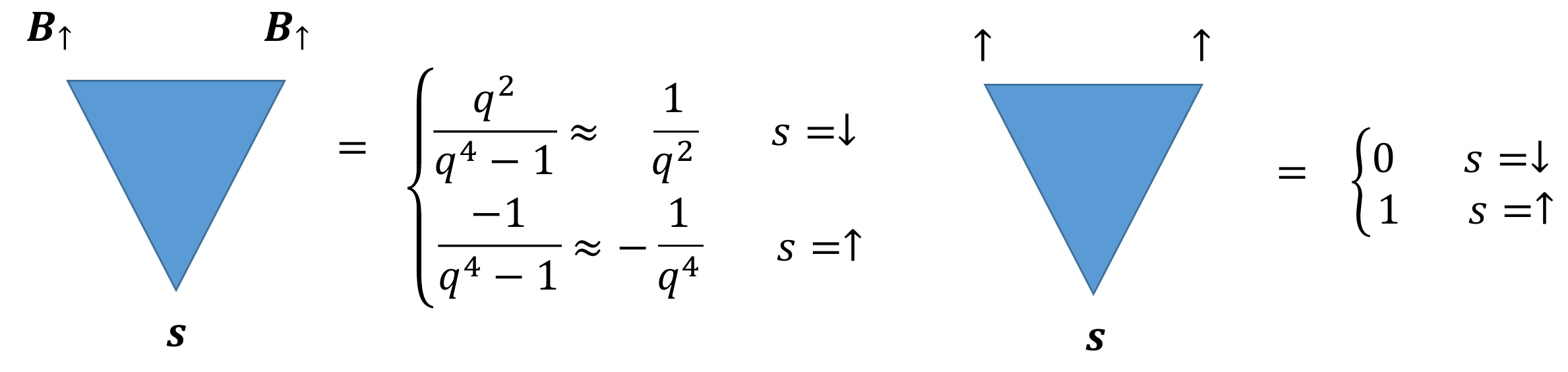}
    \caption{Rules for interactions along the top boundary.}
    \label{fig:top_rules}
\end{figure}

\item One can see that the interactions along the boundaries are the same for all nontrivial operators, so we can represent any nontrivial operator on the boundaries as $B_{\uparrow}$. The rules along the bottom and top boundaries are shown in Fig.~\ref{fig:bottom_rules} and ~\ref{fig:top_rules}.

\begin{figure}[!h]
    \centering
    \includegraphics[width=11cm]{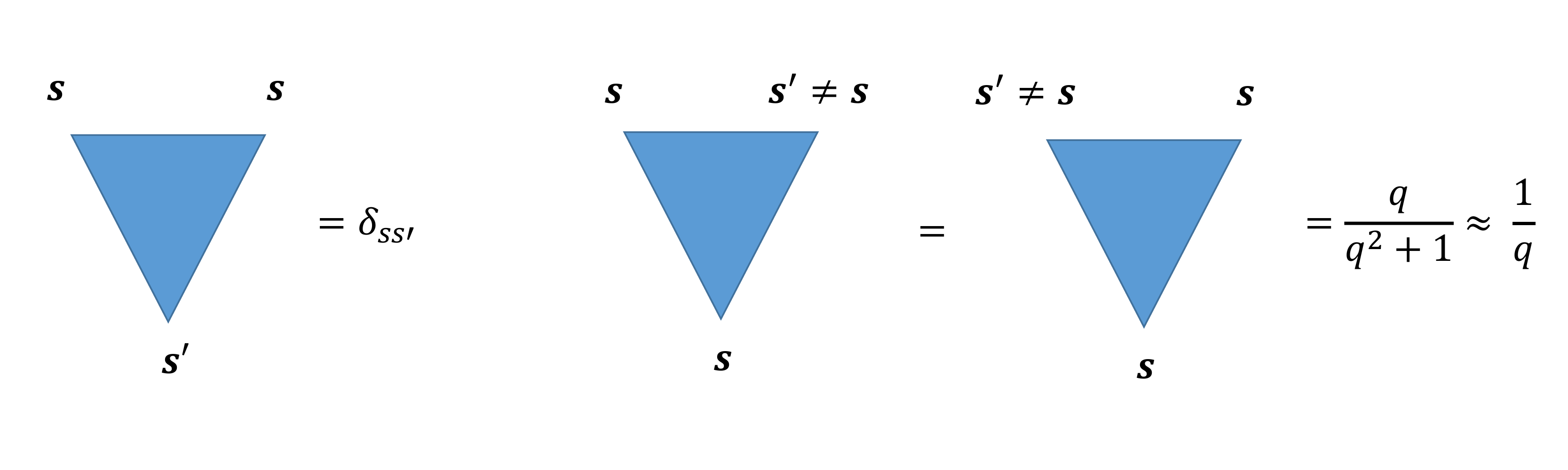}
    \caption{Interactions among the bulk spins, with $s, s'$ taking values in $\uparrow, \downarrow$. These rules will also apply to the interactions along the top boundary for observables in Appendix~\ref{app:rgd} and~\ref{app:renyi}.}
    \label{fig:bulk_rules}
\end{figure}

\item Lattice points in the bulk of the lattice correspond to locations of local unitaries in Fig.~\ref{fig:circuit_structure},
and on each lattice point lies a spin 
 taking value $\upa$ or $\doa$.  The interactions among spins in the bulk are specified by the rules of Fig.~\ref{fig:bulk_rules}.

\item The partition function is obtained by summing over all possible configurations of bulk spins, and the weight of a given configuration in the partition function is obtained
by multiplying the contributions from each interaction vertices in the bulk and along the boundaries.
\een

\subsection{Derivation of the \rgd} \label{app:rgd}

Now let us consider the probability $P_{\sO_\al}^{(A)} (t) =  \sum_{\b \; \text{with void $A$}} \coeff $ for a basis operator $\sO_\al$ to develop a void in region $A = A_1 \cdots A_n$ in random unitary circuits in the large $q$ limit. We will take $A$ to be in the future light cone of $\sO^\al$, as otherwise $P_{\sO_\al}^{(A)} (t)$ is automatically zero. 

\begin{figure}[!h]
    \centering
    \includegraphics[width=12cm]{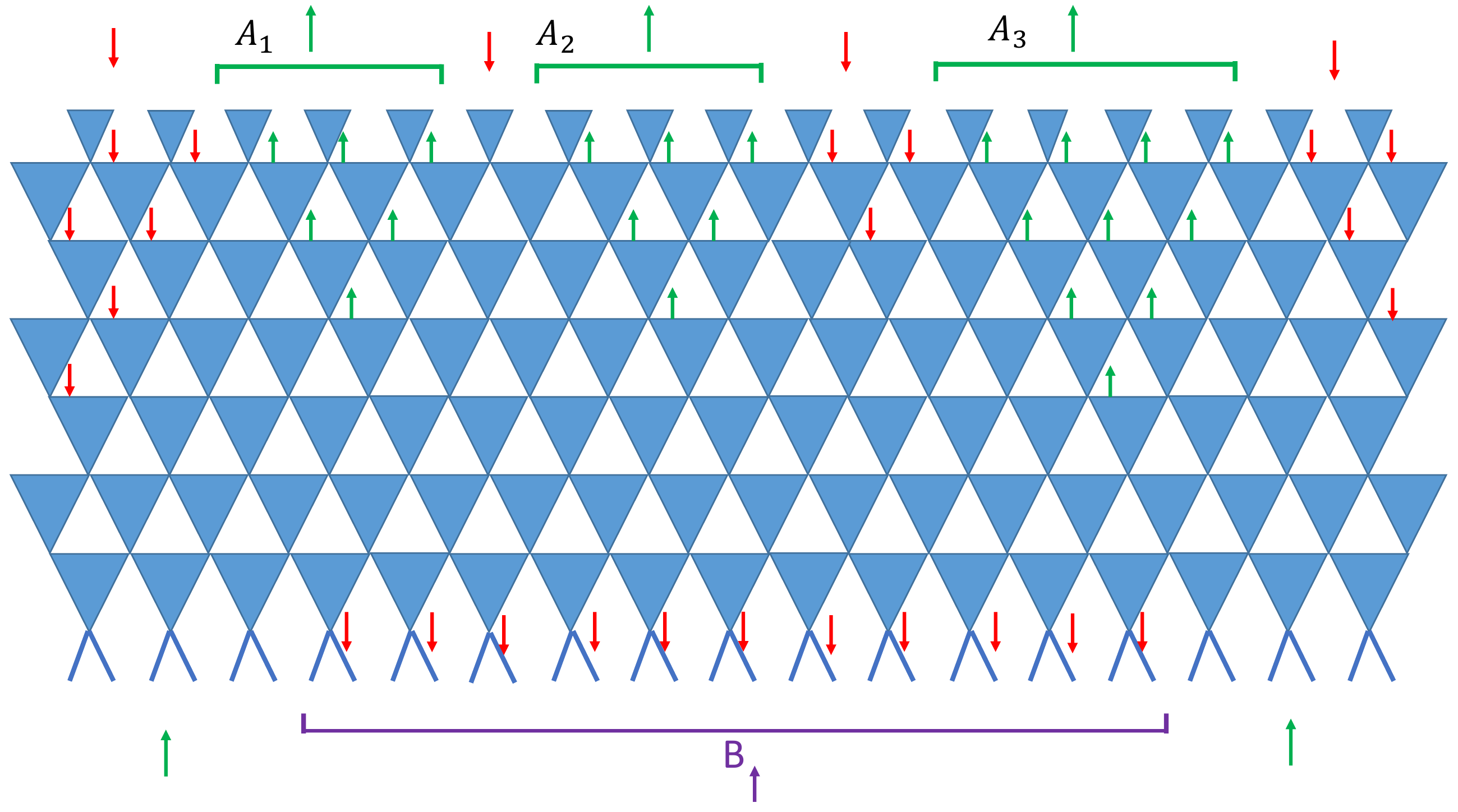}
    \caption{We show the lattice for the partition function corresponding to \eqref{P_triv} for a case with $A$ consisting of three intervals and $t=7$. The boundary conditions, and the spins in the bulk fixed by the boundary conditions through the rules in Fig.~\ref{fig:bulk_rules} and Fig.~\ref{fig:bottom_rules}, are shown explicitly. The remaining spins can be either up or down, but in the large $q$ limit we need to consider a relatively small number of configurations.}
    \label{fig:pf_lattice}
\end{figure}

To find $P_{\sO_\al}^{(A)} (t)$,  it is convenient to consider a 
slightly different quantity 
\begin{equation} \label{q_qty}
    Q_{\al}^{(\bar A)} (t)  =
     \sum_{\b \in \bar A} \coeff  \geq P_{\sO_\al}^{(A)} (t) 
\end{equation}
which also includes possible contributions from
processes in which $\sO_\al$ evolves to operators trivial in some disconnected parts of $\bar A$. Such contributions are not in $P_{\sO_\al}^{(A)} (t)$. Summing~\eqref{op_exp} over all $\sO_\b$ which have the identity 
in $A$ we find 
 \begin{equation} 
Q_{\al}^{(\bar A)} (t)= q^{|\bar{A}|} \bra{{\sO_{\alpha}}_{\uparrow}}~ (U \otimes U^{\ast}\otimes U \otimes U^{\ast})~\ket{\uparrow}_A \ket{\downarrow}_{\bar{A}} 
\label{P_triv}
\end{equation}
where $\ket{\doa_{\bar A}} = \otimes_{i \in \bar A} \ket{\downarrow}_i$ and we have used the fact that
\begin{equation}
\ket{\downarrow}_i = \frac{1}{q}\sum_{a=0}^{q^2-1} \ket{{O^i_a}_{\uparrow}} 
\label{down_spin}
\end{equation}
where the sum over $a$ runs over the complete set of basis operators at site $i$. {Now \eqref{P_triv} can be calculated with a partition function with boundary conditions as shown in Fig.~\ref{fig:pf_lattice}, with the interactions on the top boundary given by Fig.~\ref{fig:bulk_rules}.}
%We can include the prefactor $q^{|\overline{A}|}$ in \eqref{P_triv} by treating each $\ket{\downarrow}_i$ on the top boundary as being accompanied by a factor of $q$.  

\begin{figure}[!h]
    \centering
    \includegraphics[width=14cm]{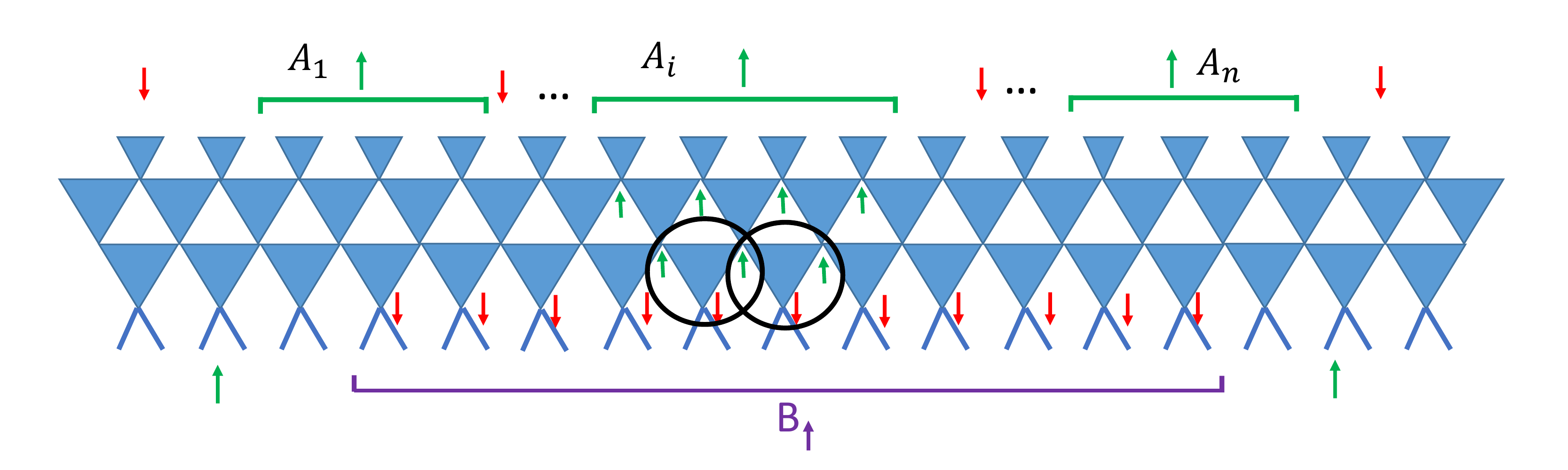}
    \caption{A case with $t<|A_i|/2$ for some $A_i$. Any configuration in the partition function at such times contributes 0 due to the circled interactions.} 
    \label{fig:small_time}
\end{figure}

We first consider a basis operator $\sO_{\alpha}$ with no initial voids. In the situation where there exists some $A_i$ for which $t < |A_i|/2$, the rules in Fig.~\ref{fig:bulk_rules} fix all spins in the past domain of dependence of each $A_i$ to be $\uparrow$ in any configuration in the partition function, while the rules in Fig.~\ref{fig:bottom_rules} fix the spins attached to $\sO_{\alpha}$ on the lower boundary to be $\downarrow$. So  we get a set of interactions like the circled interactions in Fig.~\ref{fig:small_time} along the bottom boundary, and hence the contribution from any configuration to~\eqref{P_triv} is zero, and thus 
$P_{\sO_\al}^{(A)} (t) =0$.  

We now consider the situation $t >  |A_i|/2$ for all $i$. From the rules of Fig.~\ref{fig:bulk_rules} and Fig.~\ref{fig:bottom_rules}, in the large $q$ limit, the computation of the partition function corresponding to~\eqref{P_triv} reduces to finding 
domain walls between up and down spins: for each triangle in the lattice that a domain wall passes through, we get a factor of $\frac{1}{q}$.  Starting from the top boundary, a 
domain wall can either reach the lower boundary or combine with another one to enclose intervals on the top boundary.  
A domain wall which reaches the lower boundary contributes a factor of $q^{-t}$, and the shortest domain wall which encloses an interval of length $l$  contributes $q^{-l}$ (which gives the leading contribution in the large $q$ limit).\footnote{At finite $q$, it is not sufficient to simply know the length of a domain wall, as a combinatorial factor needs to be included in each configuration to count different possible paths of a given length. But in the large $q$ limit, we can ignore this $q$-independent combinatorial factor.}
% So this schematic representation is sufficient.

\begin{figure}[!h]
    \centering
    \includegraphics[width=12cm]{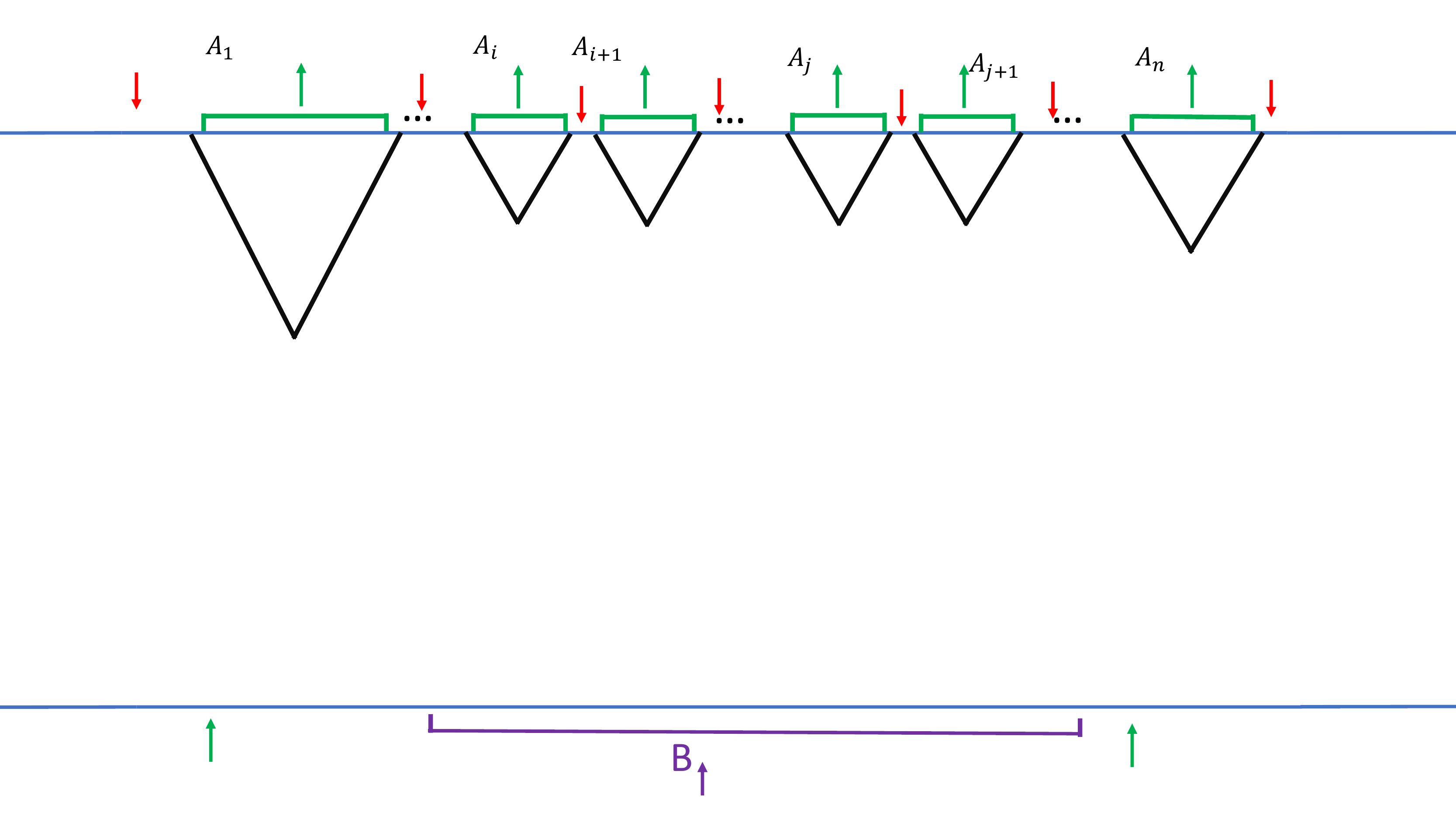}
    \caption{The configuration contributing to $P_{\sO_\al}^{(A)} (t)$. 
    In this plot to highlight the domain wall structure, we suppress the lattice in the bulk, only showing the domain walls between up and down spins, which are represented by the black lines.}  
    \label{fig:gap_config}
\end{figure}

\begin{figure}[!h]
    \centering
    \includegraphics[width=12cm]{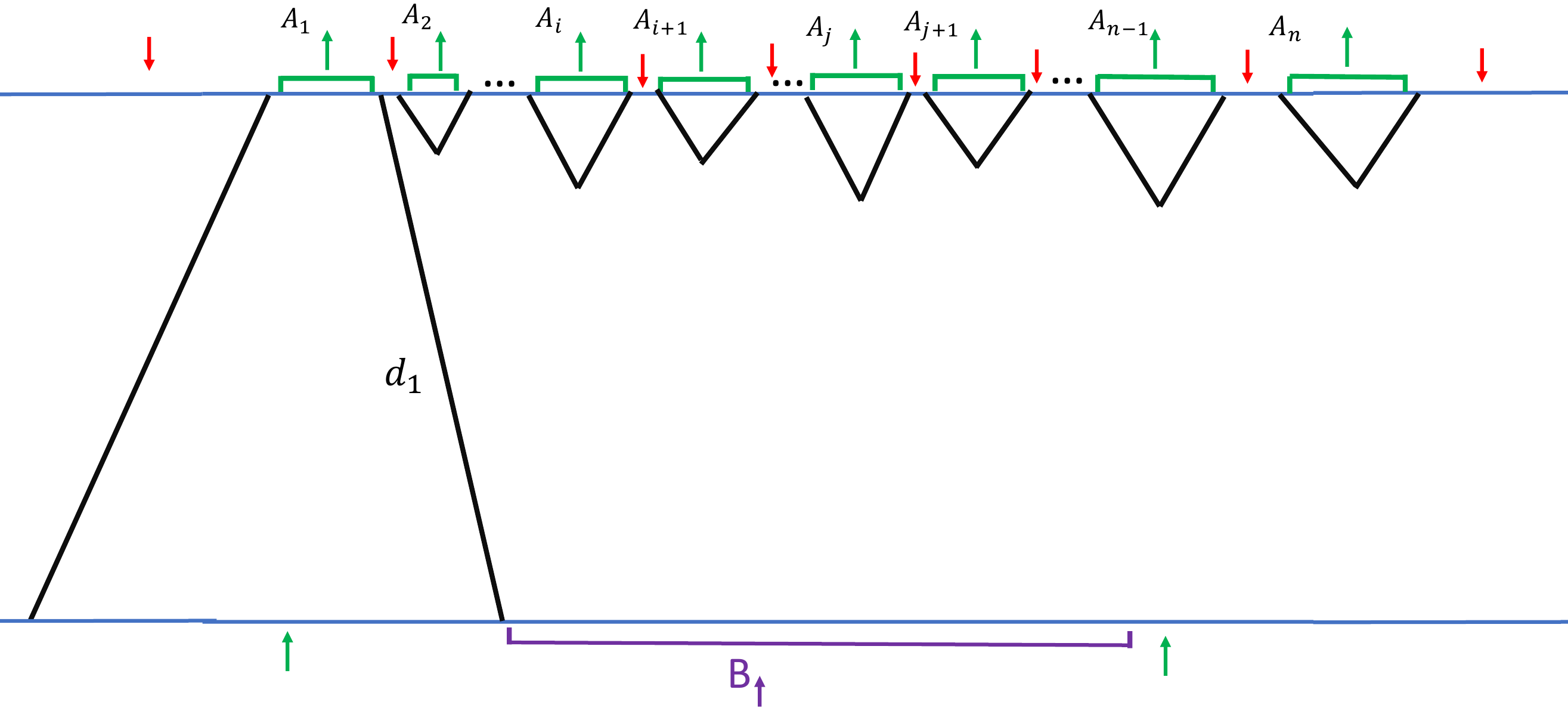}
    \caption{A domain wall configuration that contributes to \eqref{P_triv}, but not to $P_{\sO_\al}^{(A)} (t)$. It describes processes in which $\sO_\al$ evolves to operators trivial in the region to the left of $A_1$.}
    \label{fig:notgap_1}
\end{figure}

One possible configuration contributing to~\eqref{P_triv} is shown in Fig.~\ref{fig:gap_config}, with the domain walls between the up and down spins enclosing each of the $A_i$.  Since all spins on the lower most bulk layer are $\downarrow$ in this configuration, we get a factor of $q^{-L}$ from the bottom boundary. The domain walls give a factor of $q^{-(|A_1|+...+|A_n|)}$. Combing these factors with the prefactor $q^{|\bar{A}|}= q^{L-(|A_1|+...+|A_n|)}$ in~\eqref{P_triv}, we find the total contribution from this configuration is $q^{-2(|A_1|+...+|A_n|)}$. There are other possible domain wall configurations contributing to~\eqref{P_triv}, but all these other configurations correspond to the final operators in \eqref{q_qty} being trivial in some disconnected parts of $\bar A$. An example is given in Fig.~\ref{fig:notgap_1}.
Thus only the configuration of Fig.~\ref{fig:gap_config} contributes to $P_{\sO_\al}^{(A)} (t)$, and we conclude that 
\be 
P_{\sO_\al}^{(A)} (t) = q^{-2(|A_1|+...+|A_n|)}  \ . \label{rg}
\ee

\begin{figure}[!h]
\centering
    \includegraphics[width=10cm]{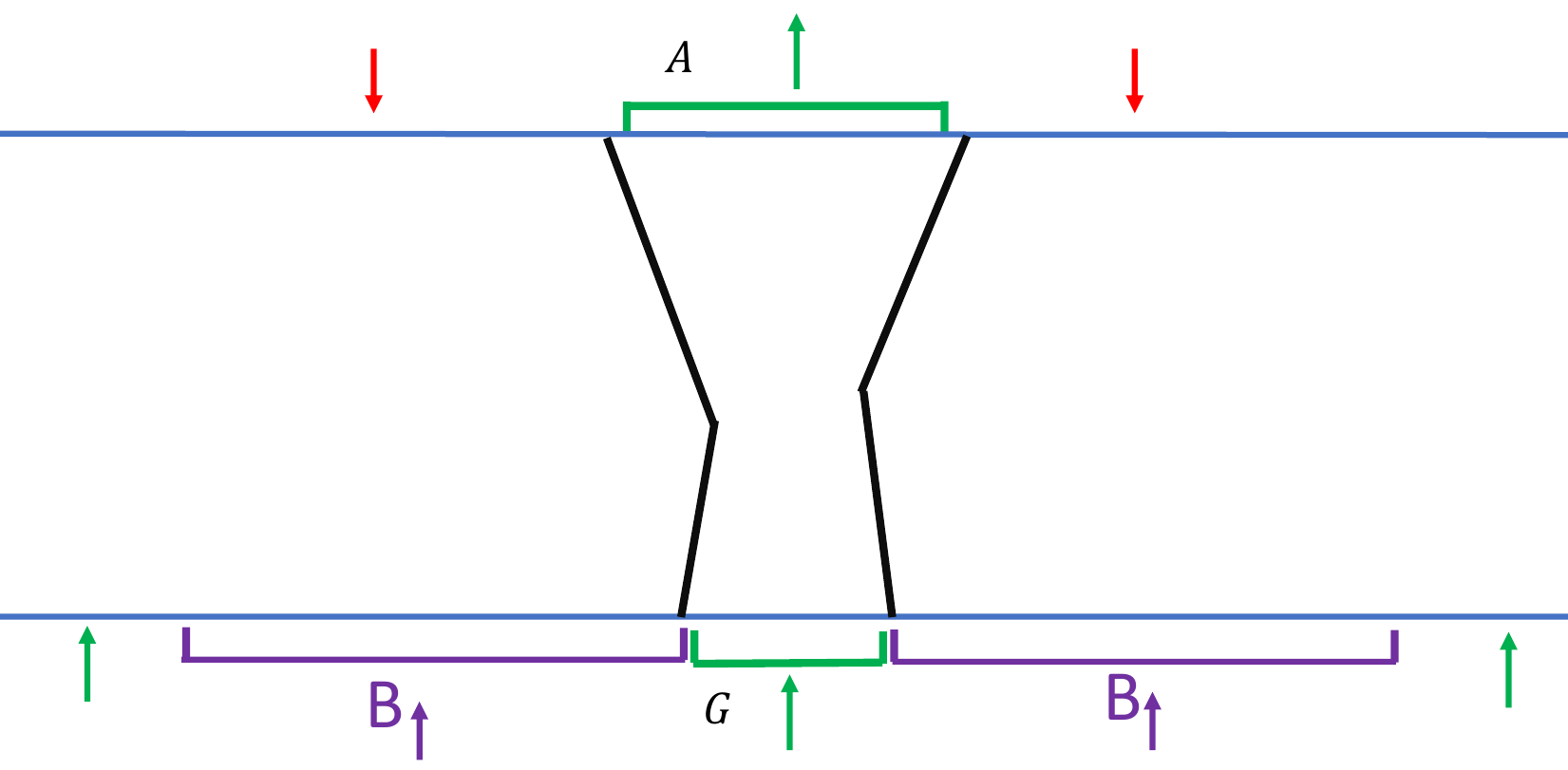}
    \caption{A domain wall configuration that contributes to $P_{\sO_\al}^{(A)} (t)$ in the case where the initial operator $\sO_{\alpha}$ has a single void in region $G$, and $A$ is a single interval. This configuration describes processes in which the disconnected parts of $\sO_\al$ evolve independently to left and right of $A$. This contribution evaluates to $q^{-|A|+|G|-2t}$, and is bigger than the contribution from Fig.~\ref{fig:gap_config} for $t<(|G|+|A|)/2$.}
    \label{fig:initial_gap}
\end{figure}

If $\sO_{\alpha}$ has an initial void, then the domain wall configuration shown in Fig. \ref{fig:gap_config} still exists, and evaluates to the same value. However, in this case, we also have the possibility that an initial void may evolve into a final void while 
the disconnected parts of the initial operator evolve independently.
 Such processes correspond to a new configuration in the partition function for $P_{\sO_\al}^{(A)} (t)$, shown in Fig.~\ref{fig:initial_gap}. Clearly when either $|A|$ or $|G|$ is larger than $2t$, this is the only process which can contribute to $P_{\sO_\al}^{(A)} (t)$. 
  When $|G|<2t$ and $|A| < 2t$, in general both possibilities 
exist and compete. The domain wall configuration of  Fig.~\ref{fig:gap_config} gives the probability that the initial void closes and opens up a new void, and we have the same value as~\eqref{rg}. The configuration of Fig.~\ref{fig:initial_gap} gives a 
contribution $q^{-|A|+|G| - 2t} = q^{-2|A|} q^{|A| + |G| -2t}$ which dominates when $|G| > 2t - |A|$.
Note that for a generic operator~\eqref{nmb}, for $\sO_\al$ with initial void $G$, we have $|a_\al|^2 \sim q^{- 2 |G|}$
and thus the overall contribution from such operators is $q^{-|A|-|G| - 2t}$ which is subdominant compared with $q^{-2 |A|}$ for 
$|A|< 2t$.

To conclude this discussion, let us note that one can find the probability for an operator of size $l_i$  to evolve 
into an operator of size $l_f$ (assuming it is allowed by causality, and both operators do not have voids) in the large $q$ limit. This probability is independent of the initial and final operators, and is given by 
\be
p_{if} (t)= q^{-l_i - l_f -2t} \ . \label{pif}
\ee
From here one finds that the probability going to all final operators with length $l_f = l_i + 2t$ allowed by causality 
is $ q^{-l_i - l_f -2t} q^{2 l_f} = q^{l_f - l_i - 2t} =1$, which leads to the sharp light-cone growth noted in ~\eqref{op_growth_1}. 
\iffalse 
\textcolor{red}{Note that the fact that a process like the one corresponding to figure \ref{fig:initial_gap} can give the dominant contribution to $P_{\sO_\al}^{(A)} (t)$ reflects the fact, which we have left implicit so far, that the sharp light cone growth in the large $q$ limit in \eqref{op_growth_1} does not imply that the subleading probability of evolving to final operators smaller than the light cone cannot give the leading contribution to some quantities. We will see an example in appendix \ref{app:alt} to explain why we are justified in ignoring the contributions for such processes to entanglement growth, for instance when we use \eqref{myu0} and \eqref{unen} to obtain \eqref{unj0} and \eqref{unj}.}
\fi

\subsection{Derivation of the time-evolution of $S_2$ in random unitary circuits with large $q$}  \label{app:renyi}

Now let us consider the evaluation of $S_2^{(A)}$,  which is given by~\eqref{ejn}, which we copy here for convenience
\be 
e^{-S_2^{(A)} (t)} =  {1 \ov q^{|A|}}  \sum_{\al \in I} \sum_{\b \in A} | c_{\al}^\b (t)|^2  \ .
\ee
Using ~\eqref{op_exp} and summing over all $\sO_\b$ which are the identity in $\bar A$ and 
all $\sO_\al$ in $I$, we find 
\be \label{enf}
e^{-S_2^{(A)} (t)} =  q^L  \vev{ \rho_{0\upa} |U \otimes U^* \otimes U \otimes U^*| \doa_A \otimes \upa_{\bar A}} \ .
\ee
In obtaining~\eqref{enf} we again used the fact that on summing over all operators in $A$, we obtain $\ket{\doa_A}$ by 
using~\eqref{down_spin}. Also recall that $\rho_0 = \otimes\prod_i \rho_i$ with each $\rho_i$ given by~\eqref{rhy}. 
Note that the boundary conditions at top boundary are reversed compared with~\eqref{P_triv}.
In the large $q$ limit, one has $e^{-\overline{S_2^{(A)}}(t)}= \overline{e^{-S_2^{(A)}(t)}}$ and 
all $S_n$ are the same~\cite{nahum2}, thus the (minus) logarithm of the partition function corresponding to~\eqref{enf} gives 
$\overline{S_2^{(A)}}(t)$ and other entanglement entropies. 
 
The structure of the lattice is the same as in Fig.~\ref{fig:pf_lattice}, but along the top boundary, we have $\uparrow$ spins in $\bar{A}$ and $\downarrow$ spins in $A$. On the lower boundary, we have $\rho_{i\uparrow}$ at each site. 
As in last subsection, the evaluation of~\eqref{enf} boils down to summing over domain wall configurations. From the fact that each $\rho_{i}$ is a projector onto a single state in the one-site Hilbert space, we get a factor of $\frac{1}{q}$ from each site on the lower boundary, irrespective of whether the lowermost bulk layer has an $\uparrow$ or $\downarrow$ spin at that site.\footnote{If $\rho_i = \ket{\psi}\bra{\psi}$, then $\braket{abcd|\rho_{i\uparrow}}= \psi_a \psi_b^{\ast}\psi_c \psi_d^{\ast}$, so $\braket{\uparrow |\rho_{i\uparrow}}= \braket{\downarrow |\rho_{i\uparrow}}= \frac{1}{q}|\psi|^2|\psi|^2 =1/q$.} Thus irrespective of where the domain walls end, we get a factor of $\frac{1}{q^L}$ from the lower boundary, cancelling with the prefactor $q^L$ in \eqref{enf}. Effectively the bottom boundary does not play any role. 

Now consider a region $A = A_1  \cdots  A_n$ consisting of $n$ intervals $A_1 = [l_1, r_1]$, ..., $A_n=[l_n, r_n]$, separated by intervals $R_1, ..., R_{n-1}$. Due to the boundary conditions, in each non-zero configuration in the partition function, we have $2n$ starting points of domain walls on the top boundary, from each of the $l_i$ and $r_i$. As discussed in last subsection, in the large $q$ limit,  we only need to specify whether each domain wall starting on the top boundary reaches the bottom boundary, in which case we get a factor of $q^{-t}$, or joins with another domain wall to enclose an interval of length $l$, in which case we get a factor of $q^{-l}$. The latter possibility only exists for $t>l/2$. 

\begin{figure}[!h]
    \centering
    \includegraphics[width=10cm]{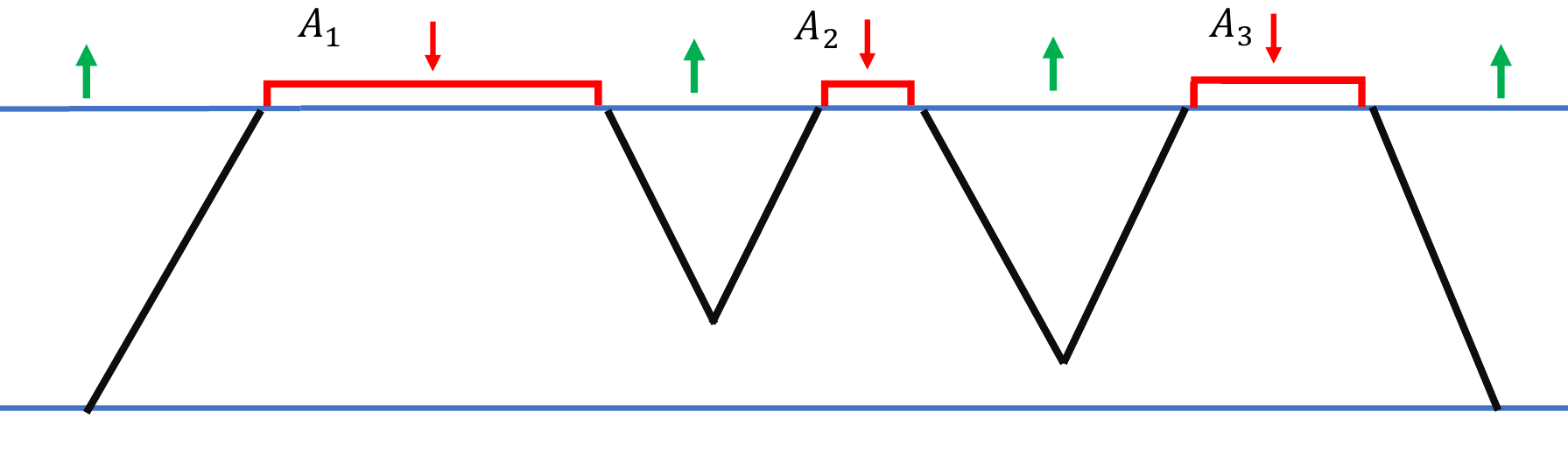}
    \caption{Example of a configuration contributing to the partition function for \eqref{enf}. The set $\gamma$ corresponding to this configuration is $\{l_1, r_3, \{l_2, r_1 \}, \{l_3, r_2\}\}$.}
    \label{fig:example_S2}
\end{figure}

Due to the boundary conditions, in order to separate regions of opposite spins, domain walls starting from some left-endpoint $l_i$ can only join with domain walls starting from some right-endpoint $r_j$, and from the rules of Fig.~\ref{fig:bulk_rules}, domain walls cannot intersect.
Let us call a pair of adjacent endpoints $l_i$ and $r_j$ an ``allowed pair" at time $t$ if $|l_i-r_j|/2\leq t$. It is convenient to  define a set $\{\gamma\}(t)$ of possible domain wall configurations. An element $\ga$ of the set  
contains some number of ``allowed pairs" $\{l_i, r_j\}$ and some number of unpaired points, such that each endpoint appears exactly once.  Paired points correspond to domain walls enclosing the region between them, while unpaired points correspond to 
domain walls reaching the bottom boundary. See Fig.~\ref{fig:example_S2} for an example. The contribution from a $\ga$-configuration evaluates to  $(\frac{1}{q})^{f_{\gamma}}$, where 
\be
f_{\gamma} =  n_{\gamma} t + \sum_{\{l_i, r_j\}\in \gamma} \mid l_i - r_j\mid \ ,
\ee
and $n_{\gamma}$ is the number of unpaired points in $\gamma$. In the large $q$ limit, the configuration with minimal $f(\ga)$ dominates.\footnote{In the large $q$ limit we can also ignore any $q$-independent prefactor, as it contributes an $O(1)$ term to $S_2^{(A)}$ after taking the logarithm.} 
 We thus find 
\begin{equation}
S^{(A)}_2(t) %= -\log(\big(\frac{1}{q}\big)^{\min_{\gamma \in \{\gamma\}(t)}f_{\gamma}}) \\
= \seq \min_{\gamma} f_\ga, \quad  s_{\text{eq}} = \log q \ .
\end{equation}

\section{Entanglement entropy from the \rgd} \label{app:alt}

Here we prove that~\eqref{hol2} follows using only the following rules: (i) \slcg, (ii) the \rgd~\eqref{rgd1}, (iii) large $q$,
for $A$ consisting of an arbitrary number of disjoint intervals. 
Considering~\eqref{ejn}, we can write $N_A (t)$ more explicitly as 
\be \label{jgh}
N_A (t) =  N (A, D(Q); t), \quad Q = A_1 R_1 \cdots R_{n-1} A_n 
\ee
Before presenting the proof, let us first note the two key elements which are heavily used:

\ben

\begin{figure}[!h]
    \centering
    \includegraphics[width=14cm]{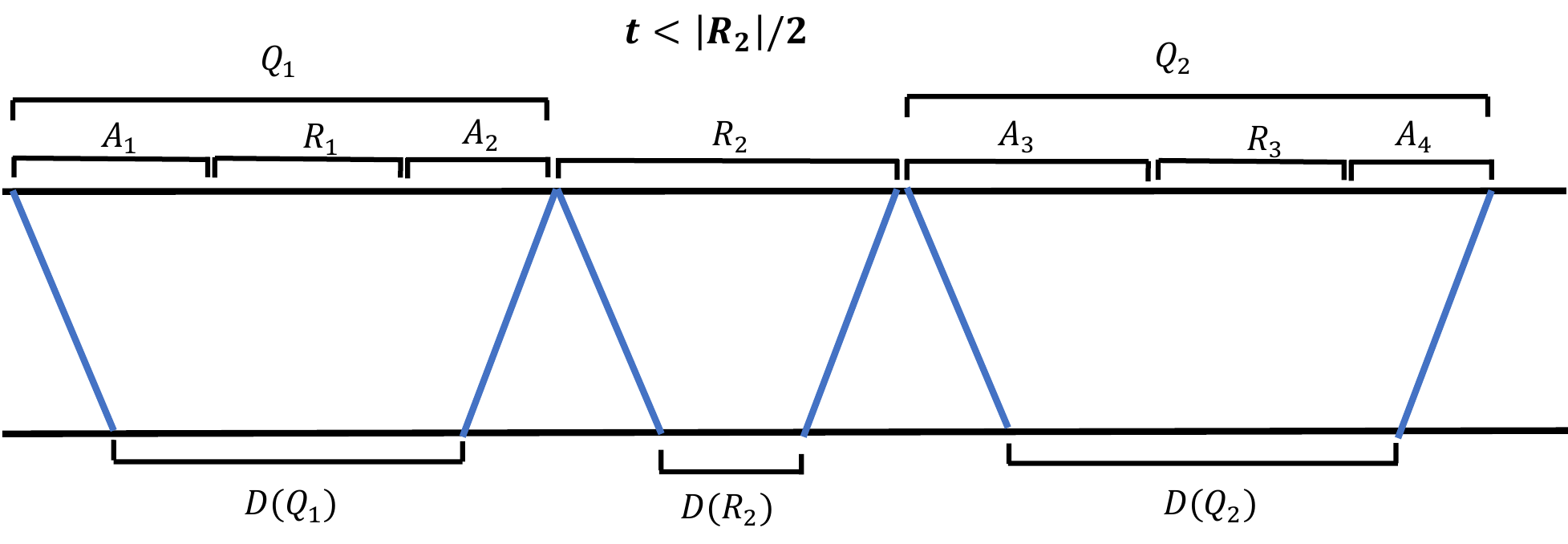}
    \caption{$R_2$ separates $A$ into two parts with $M_1 = A_1 A_2$ and $M_2 =A_3 A_4$, which can be independently considered. The initial operator must have identity in region $D(R_2)$ in order to contribute to~\eqref{jgh}, and thus the contributions from two sides of region $D(R_2)$ factorize.}
    \label{fig:er}
\end{figure}

\item At any given time $t$, if there exists some $|R_i| > 2t$,  the two parts of $A$ separated by interval $R_i$ can be 
independently considered. See Fig.~\ref{fig:er} for an example, where we have a factorized form 
\be\label{bav}
N (A, D(Q); t) = N (M_1, D(Q_1); t) N (M_2, D(Q_2); t)  \ .
\ee

\begin{figure}[!h]
\centering 
\includegraphics[width=15cm]{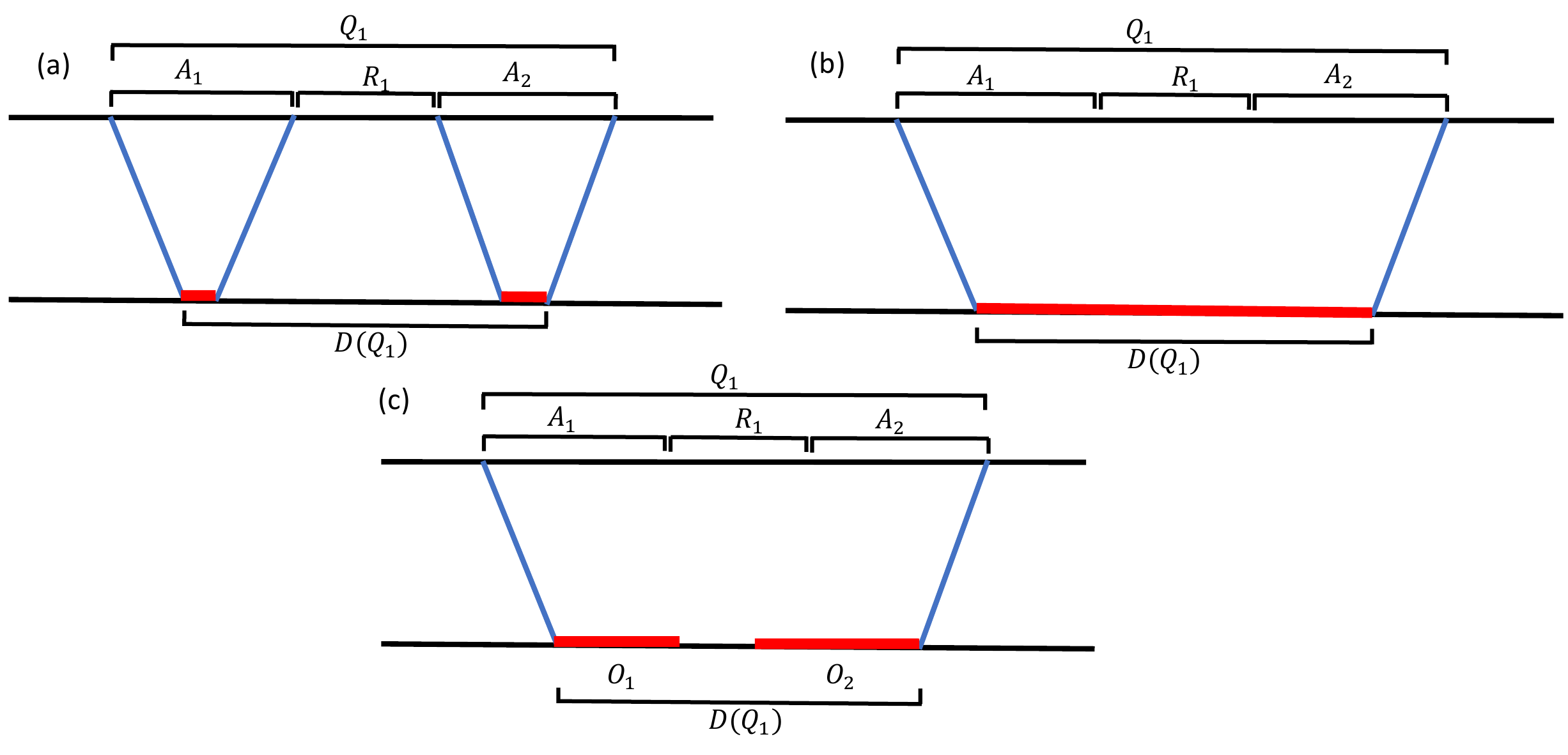}
\caption{Possible contributions to $N (A_1 A_2, D(Q_1); t)$ of Fig.~\ref{fig:er}. 
In the figure, red shaded regions have non-trivial operators at all sites, while the unshaded part of the initial operator is trivial.}
\label{fig:two_int_partitions}
\end{figure}

\item Consider one of the factorized parts, in which $2t$ is greater than all $R_i$ in that part,  for example, $Q_1$ in Fig.~\ref{fig:er}. There are possible multiple competing contributions to $N (A_1 A_2, D(Q_1); t)$, which are exhibited 
in Fig.~\ref{fig:two_int_partitions}. One contribution comes from operators which are nontrivial at all sites of $D(Q_1)$, see Fig.~\ref{fig:two_int_partitions}(b), 
which from~\eqref{rgd1} gives 
\begin{equation}
    N_{\text{(connected)}}^A(t) = M_{D(Q_1)}(t) q^{-2|R_1|}=  q^{|A_1|+ |A_2|-|R_1|-2t}
    \label{conn}
\end{equation}
where $M_{D(Q_1)}(t) = q^{|D(Q_1)|} = q^{|A_1|+ |A_2|+|R_1|-2t}$ is the number of initial basis operators in $D(Q_1)$. 
There is also a disconnected contribution from Fig.~\ref{fig:two_int_partitions}(a), where 
nontrivial operators  in $D(A_1)$ and $D(A_2)$ separated by an initial void evolve independently to region $A_1$ and $A_2$ respectively.  Note that one may also consider initial operators with a void like in Fig.~\ref{fig:two_int_partitions}(c), where the non-trivial parts of the operator are not contained within $D(A_1)$ and $D(A_2)$. Assuming sharp light-cone growth, such an operator cannot give a disconnected contribution  in which $\sO_1, \sO_2$ evolve independently to $A_1$ and $A_2$. However, such an operator can give a connected contribution, which corresponds to  situations where the initial void closes and then opens an new void. This contribution is suppressed compared to~\eqref{conn}, as the phase space for initial operators with a void is suppressed, while the probability for an individual operator to develop a void is the same. 
\een

We will prove~\eqref{hol2} by induction. For two intervals, $n=2$, we already showed this explicitly in Sec.~\ref{sec:rgd2}~(the discussion leading to~\eqref{unj0}--\eqref{unj}). Here we give another derivation which connects more directly to the form~\eqref{hol2}. From item 1 above, for $t<|R|/2$, 
\begin{equation} \label{disc}
    N(A, D(Q), t) = N_{A_1}(t) N_{A_2}(t) \quad \to \quad S_2^{(A)} =  S_2^{(A_1)}  +  S_2^{(A_2)}  \ .
\end{equation}
For $t>|R|/2$, from item 2 we have two competing contributions: the connected one, which is given by~\eqref{conn}, and the disconnected one, given by~\eqref{disc}. Comparing \eqref{conn} and \eqref{disc} in the large $q$ limit leads to the two-interval result from~\eqref{hol2}, with the connected contribution~\eqref{conn} corresponding to having the pair $(r_1, l_2)$ in the endpoint configuration $\gamma$.

Assuming we have~\eqref{hol2} up to $n=k$ intervals, let us now consider $n=k +1$. 

 Let us denote the collection of $R_i$'s which are greater than $2t$ by $\sR_{\rm big}$. 
Such $R_i$'s generate a partition $\sP(t)$ of $A$ into $m$ parts (where $m$ can range from $1$ to $n$ depending on the time), $M_1, ..., M_{m}$. For the reason stated in Fig.~\ref{fig:er} and item 1 above, the contribution from each $M_i$ factorizes, and we thus have 
\be \label{pm}
S_2^{(A)} (t) = \sum_{j=1}^m S_2^{(M_j)} (t)  \ .
\ee
The same thing happens to~\eqref{hol2}: if there exists a $|R_i| > 2t$, then the two parts of $A$ separated by interval $R_i$ can be independently minimized as there are no allowed pairing between end points on the left of $R_i$ and those on the right of $R_i$. Thus equation~\eqref{pm} also applies to~\eqref{hol2}, and each term in the sum of~\eqref{pm} is given by~\eqref{hol2} from our assumption regarding $n \leq k$.

\begin{figure}[!h]
    \centering
    \includegraphics[width=11cm]{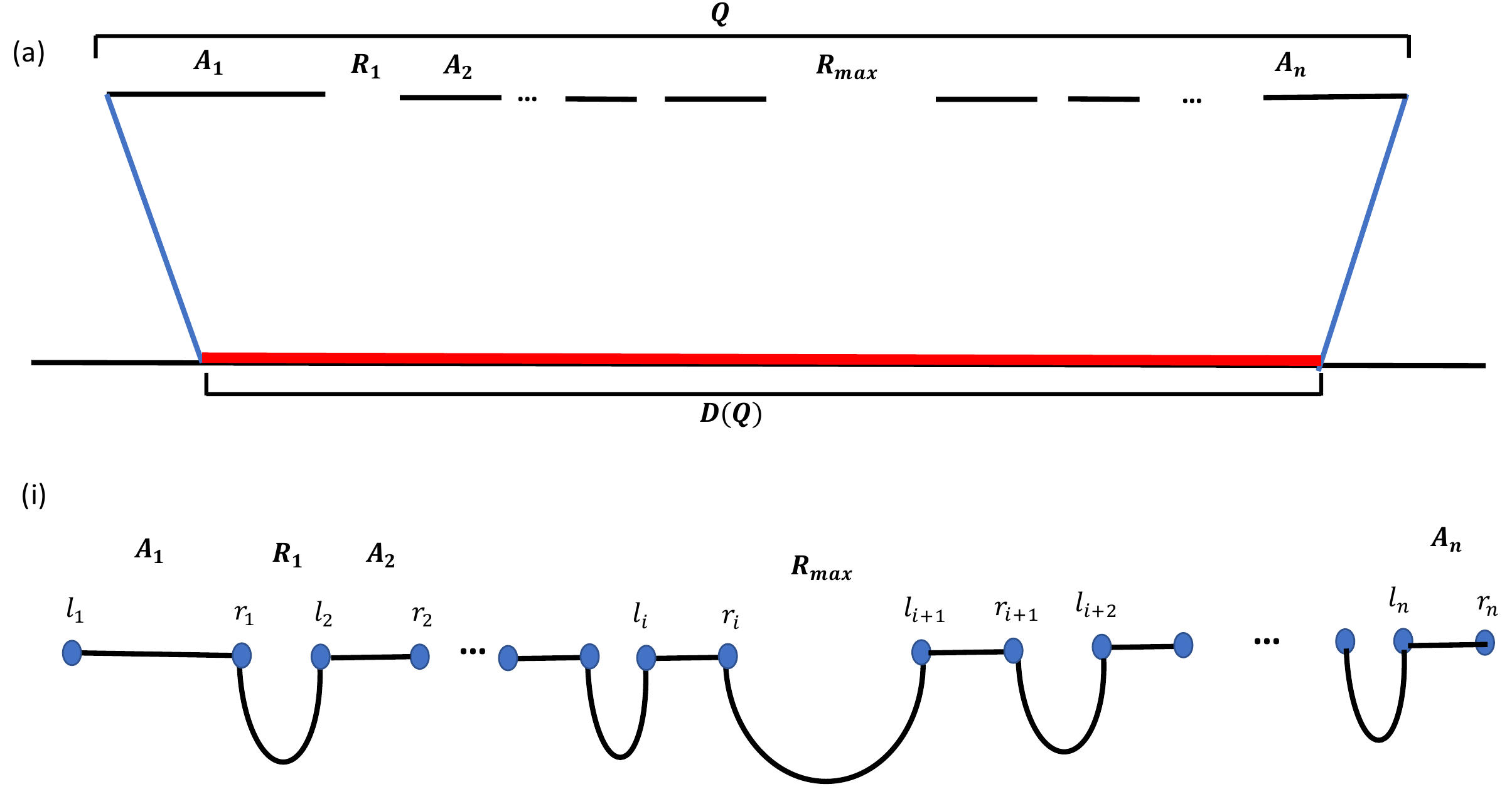}
    \includegraphics[width=11cm]{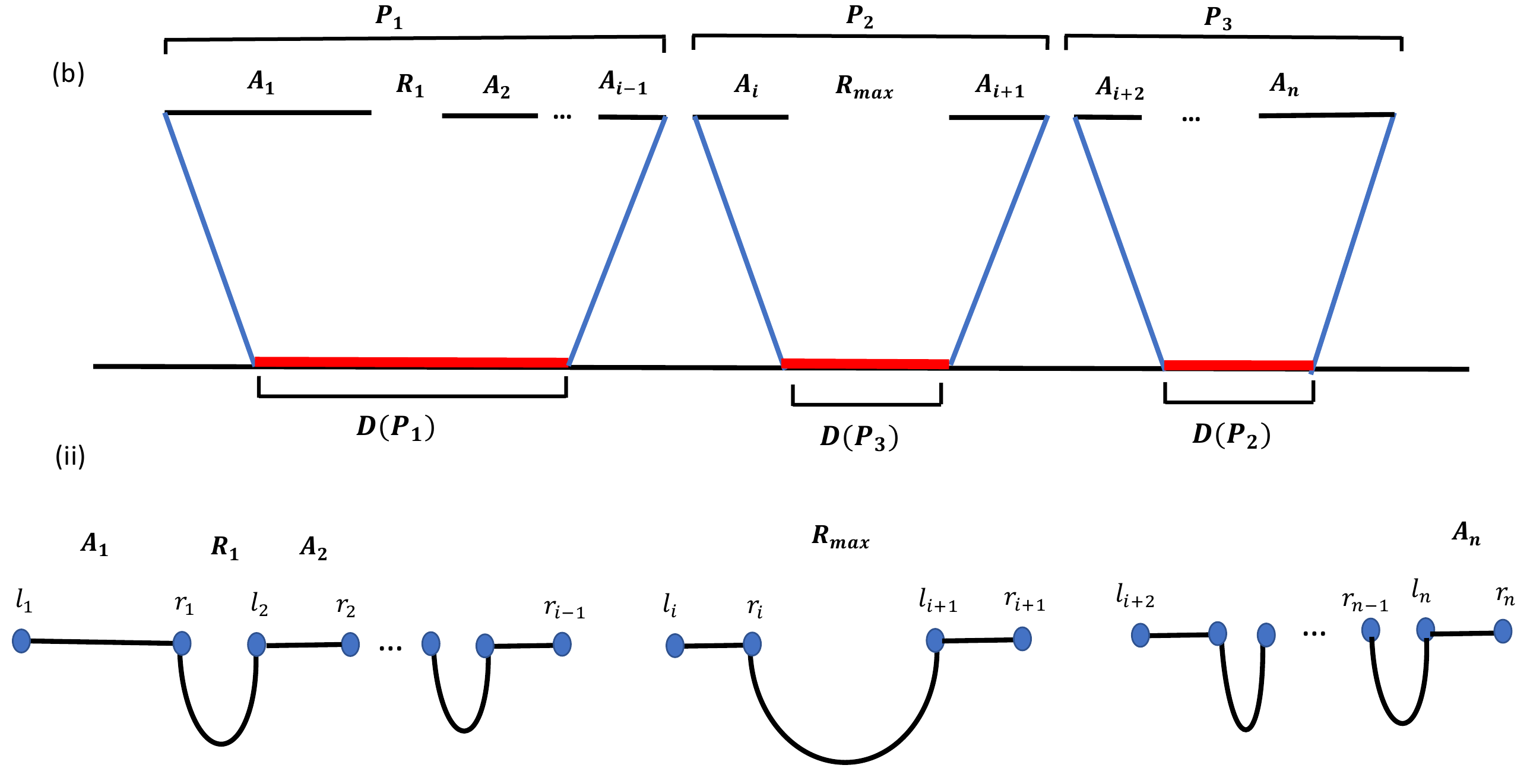}
    \caption{Two sets of initial operators contributing to \eqref{ejn}, and the corresponding endpoint configurations $\gamma$.  In (a), initial operators nontrivial everywhere in $D(Q)$ develop a void in $R \equiv R_1 \cdots R_{n-1}$, 
 contributing $q^{|Q|-2t} q^{-2 |R|}= q^{|A| +|R| - 2t} q^{-2 |R|} = q^{|A|-|R|-2t}$ to $N_A (t)$, leading to a possible contribution $S^{(A)}_2 (t) = \seq (|R| + 2t)$. We get the same contribution from $\gamma$ shown in (i), where $l_1$ and $r_n$ are the only unpaired points, and all $r_m$ for $m<n$ are paired with $l_{m+1}$. In (b), initial operators nontrivial everywhere in each $D(P_i)$ go into $P_i$, forming voids in the $R_j$'s contained within the $P_i$'s. This contributes
  $q^{|P_1|-2t} q^{-2(|R_1|+...+|R_{i-2}|)}q^{|P_2|-2t} q^{-2|R_{\text{max}}|}q^{|P_3|-2t} q^{-2(|R_{i+2}|+...+|R_{n-1}|)}$ to $N_A (t)$, leading to a possible contribution $S^{(A)}_2 (t) = \seq (|R|-|R_{i-1}|-|R_{i+1}| + 6t)$. We get the same contribution from $\gamma$ shown in (ii), where $l_1$, $r_{i-1}$, $l_i$, $r_{i+1}$, $l_{i+2}$, and $r_n$ are the only unpaired points, and all $r_m$ for $m \neq i-1, i+1, n$ are paired with $l_{m+1}$.  }
    \label{fig:er1}
\end{figure}

We increase $t$ until $t = r/2$ when the set $\sR_{\rm big}$ becomes empty, where $r$ is size of the largest $R_i$, which we call $R_{\rm max}$. Slightly before reaching $t=r/2$, from the \rgd~\eqref{rgd1} initial operators cannot develop a void in region $R_{\rm max}$, which in the language of~\eqref{hol2} corresponds to the fact that the two end points of $R_{\rm max}$ 
cannot be paired with each other. Now when $t > r/2$, from the \rgd~\eqref{rgd1}, there are new contributions coming from 
initial operators developing a void in region $R_{\rm max}$, which compete with previously existing ones. Note the new contributions include the connected one for the full region $D(Q)$, but also disconnected ones. See Fig.~\ref{fig:er1}  for two examples. These new contributions are in one-to-one correspondence 
with new $\ga$-configurations in~\eqref{hol2} which come from pairing the two end points of $R_{\rm max}$. 
One can also readily check that their respective contributions agree. 
%See Fig.~\ref{fig:er1}  for two examples of contributions from different sets of initial intervals, and the corresponding endpoint configurations. %As we justified in detail in the $n=2$ case, it is sufficient to consider initial operators which are non-trivial everywhere in the domains of dependence of $A_iR_i...A_j$ for different choices of $A_i$,$A_j$,  for which we can assume the random void distribution. 
Thus concludes the proof.

In the above proof, we assumed the random void distribution for all initial operators, an assumption that is not precisely true in random unitary circuits in the large $q$ limit due to the subtlety noted at the end of Appendix~\ref{app:rgd}. Yet we found by the partition function calculation of Appendix \ref{app:renyi} that equation \eqref{hol2} is true in random unitary circuits in the large $q$ limit, indicating that the cases where $P_{\sO_\al}^{(A)} (t)$ for initial operators with voids is not given by the random void distribution can be ignored in the calculation of $S_2$. For instance, this means that in  Fig.~\ref{fig:two_int_partitions}(c), the cases in random unitary circuits where the disconnected contribution (due to independent evolution of $O_1$ and $O_2$ respectively into $A_1$ and $A_2$) is dominant can be ignored. Indeed, on using~\eqref{pif} and counting the number of relevant initial and final operators, we can check that the collective disconnected contribution from all initial operators like in  Fig.~\ref{fig:two_int_partitions}(c) is the same as that from Fig.~\ref{fig:two_int_partitions}(a), changing $S_2$ at most by a $q$-independent $O(1)$ term which can be neglected in the large $q$ limit. 
%So we can directly assume sharp light-cone growth and the random void distribution, which immediately reduces the comparison between cases (a), (b) and (c) to a comparison between cases  (a) and (b).}

\end{document}